\documentclass[aps,prx,amsmath,amssymb,footinbib,showpacs,twocolumn,superscriptaddress]{revtex4-1}
\usepackage{amsmath}
\usepackage{amssymb}
\usepackage{amsthm}
\usepackage{setspace}
\usepackage{graphicx}
\usepackage{braket}
\usepackage{mathrsfs}
\usepackage{physics}
\usepackage{float}
\usepackage[colorlinks = true,linkcolor = red,urlcolor  = blue,citecolor = blue,anchorcolor = blue]{hyperref}
\usepackage[utf8]{inputenc}
\usepackage[english]{babel}

\newcommand{\angstrom}{\text{\normalfont\AA}}
\begin{document}
\title{Metamorphosis of Andreev bound states into Majorana bound states in pristine nanowires}
\author{Yingyi Huang}
\affiliation{
Condensed Matter Theory Center and Joint Quantum Institute, Department of Physics, University of Maryland, College Park, MD 20742, USA}
\affiliation{
State Key Laboratory of Optoelectronic Materials and Technologies, School of Physics, Sun Yat-sen University, Guangzhou 510275, China}
\author{Haining Pan}
\affiliation{
Condensed Matter Theory Center and Joint Quantum Institute, Department of Physics, University of Maryland, College Park, MD 20742, USA}
\author{Chun-Xiao Liu}
\affiliation{
Condensed Matter Theory Center and Joint Quantum Institute, Department of Physics, University of Maryland, College Park, MD 20742, USA}
\author{Jay D. Sau}
\affiliation{
Condensed Matter Theory Center and Joint Quantum Institute, Department of Physics, University of Maryland, College Park, MD 20742, USA}
\author{Tudor D. Stanescu}
\affiliation{
Condensed Matter Theory Center and Joint Quantum Institute, Department of Physics, University of Maryland, College Park, MD 20742, USA}
\affiliation{
Department of Physics and Astronomy, West Virginia University, Morgantown, WV 26506}
\author{S. Das Sarma}
\affiliation{
Condensed Matter Theory Center and Joint Quantum Institute, Department of Physics, University of Maryland, College Park, MD 20742, USA}

\begin{abstract}
We show theoretically that in the generic finite chemical potential situation, the clean superconducting spin-orbit-coupled nanowire has two distinct nontopological regimes as a function of Zeeman splitting (below the topological quantum phase transition): one is characterized by finite-energy in-gap Andreev bound states, while the other has only extended bulk states. The Andreev bound state regime is characterized by strong features in the tunneling spectra creating a ``gap closure" signature, but no ``gap reopening" signature should be apparent above the topological quantum phase transition, in agreement with most recent experimental observations.  The gap closure feature is actually the coming together of the Andreev bound states at high chemical potential rather than a simple trivial gap of extended bulk states closing at the transition. Our theoretical finding establishes the generic intrinsic Andreev bound states on the trivial side of the topological quantum phase transition as the main contributors to the tunneling conductance spectra, providing a generic interpretation of existing experiments in clean Majorana nanowires. Our work also explains why experimental tunnel conductance spectra generically have gap closing features below the topological quantum phase transition, but no gap opening features above it.
\end{abstract}
\date{\rm\today}
\maketitle

\section{introduction}\label{sec:intro}
Interest in the solid-state manifestations of non-Abelian Majorana bound states (MBS) and the consequent prospects for topological quantum computation have led to intensive theoretical and experimental studies~\cite{Nayak2008Non-Abelian, DasSarma2015Majorana,Alicea2012New,Elliott2015Colloquium,Stanescu2013Majorana,Leijnse2012Introduction,Beenakker2013Search,Lutchyn2017Realizing,jiang2013non,sato2016majorana,Sato2016Topological,Aguado2017Majorana} of strongly spin-orbit-coupled semiconductor nanowires (InSb or InAs) in proximity to ordinary $s$-wave superconductors (Al or Nb) in an applied magnetic field, following specific theoretical predictions~\cite{Sau2010Generic,Lutchyn2010Majorana,Oreg2010Helical,Sau2010Non} that such semiconductor-superconductor hybrid structures should produce artificially induced topological superconductivity (TSC). In this particular context, this is essentially a spinless $ p $-wave one-dimensional (1D) superconductor~\cite{Kitaev2001Unpaired} produced in the nanowire by the combination of spin-orbit coupling, Zeeman spin splitting, and $ s $-wave proximity superconductivity.  Experiments~\cite{Mourik2012Signatures,Das2012Zero,Deng2012Anomalous,Churchill2013Superconductor,Finck2013Anomalous,Albrecht2016Exponential,Chen2016Experimental,Zhang2016Ballistic,Deng2016Majorana,Zhang2017Quantized,Nichele2017Scaling} have indeed observed the predicted~\cite{Sengupta2001Midgap, Law2009Majorana} zero-bias conductance peaks (ZBCP) associated with the topological MBS (in fact, even the predicted ZBCP quantization has recently been observed~\cite{Zhang2017Quantized}), but whether the ZBCP actually arises from MBS or not has remained controversial because trivial ZBCP could arise from ordinary (i.e., nontopological) subgap Andreev bound states (ABS) due to chemical potential fluctuations in the nanowire, which cannot be ruled out in realistic systems~\cite{Brouwer2011Topological,Chiu2017Conductance,Liu2017Andreev,Moore2016Majorana}. The problem of accidental ABS-induced trivial ZBCP mimicking MBS-induced topological ZBCP in realistic nanowires is the most formidable current challenge in the field, and many theoretical suggestions have recently been made~\cite{Liu2017Andreev,Prada2017Measuring,Clarke2017Experimentally,Liu2018Distinguishing,Chiu2017Interference,Liu2018Measuring} with a goal toward discerning between ABS and MBS in nanowires. However, the experimental situation remains murky, with no clear consensus on whether the experimentally observed ZBCP arise from topological MBS or trivial ABS.

The current theoretical work is also on ABS in nanowires, but arises from fundamentally different physics compared with the accidental ABS produced by chemical potential fluctuations which have been much discussed in the recent literature. The ABS in the current work is \emph{intrinsic} and happens in pristine nanowires with no disorder whatsoever (i.e., no quantum dots or chemical potential variations), and is generically present below the topological quantum phase transition (TQPT) (i.e., for Zeeman splitting $V_Z<V_{Zc}$, where $V_{Zc}=\sqrt{\Delta^2 + \mu ^2}$ with $\Delta$, $\mu$ being, respectively, the proximity-induced SC gap and chemical potential in the nanowire and $V_{Zc}$ the critical spin splitting necessary for producing TSC~\cite{Sau2010Generic}), as long as the chemical potential ($\mu$) in the system is finite. In fact, we find a characteristic Zeeman field $V_{Zt}<V_{Zc}$ below which (i.e., for $0<V_{Z}<V_{Zt}<V_{Zc}$) the \textit{intrinsic} nanowire states defining the SC proximity gap are localized ABS rather than extended bulk band states. The generic existence of $V_{Zt}$ in pristine nanowires for finite $\mu$ and the associated nanowire SC gap being an effective ABS gap on the nontopological side of the TQPT are the results being presented in this work.  We note that for large $\mu$, $V_{Zt}$ could be arbitrarily close to $V_{Zc}$ whereas for $\mu=0$, $V_{Zt}=0$ and the ABS no longer exists.  We therefore conclude that for arbitrary $\mu$, it is likely that the experimentally observed gap closing in pristine nanowires is simply the coming together of the ABS found in this work, and not the closing of the bulk SC gap. This result may explain why the best current nanowire Majorana experiments typically observe strong conductance features associated with gap closing, but none for gap reopening.

The paper is organized as follows. In Sec.~\ref{sec:model}, we give the effective model for the Majorana nanowire and introduce the numerical method for the tunneling conductance calculation. In Sec.~\ref{sec:numerical}, the numerical results of both differential conductance and energy spectra are presented. In Sec.~\ref{sec:phase}, we show the wave functions of the two lowest-energy eigenstates and give a phase diagram for the Majorana nanowire. We then give in Sec.~\ref{sec:Analytical} the analytical proof for the existence of the intrinsic ABS and derive the analytic form of $V_{Zt}$. In Sec.~\ref{sec:shorter}, we discuss experimental implications of our results providing additional results for shorter wires used in the current laboratory samples. In Sec.~\ref{sec:QD}, we go beyond the pristine clean nanowires and investigate nanowires in the presence of an external quantum dot discussing the interplay of intrinsic and extrinsic Andreev bound states. We conclude in Sec.~\ref{sec:conclusion}. Additional simulation results for both the intrinsic and extrinsic Andreev bound states are provided in the Appendices for completeness. All results except those in Sec.~\ref{sec:QD} presented in this paper are obtained at zero temperature and zero disorder in pristine systems, whereas the results in Sec.~\ref{sec:QD} have the additional quantum dot potential in order to create the extrinsic ABS.

\section{effective model and numerical method}\label{sec:model}
\begin{figure*}
\raggedleft
\includegraphics[width=1.0\textwidth]{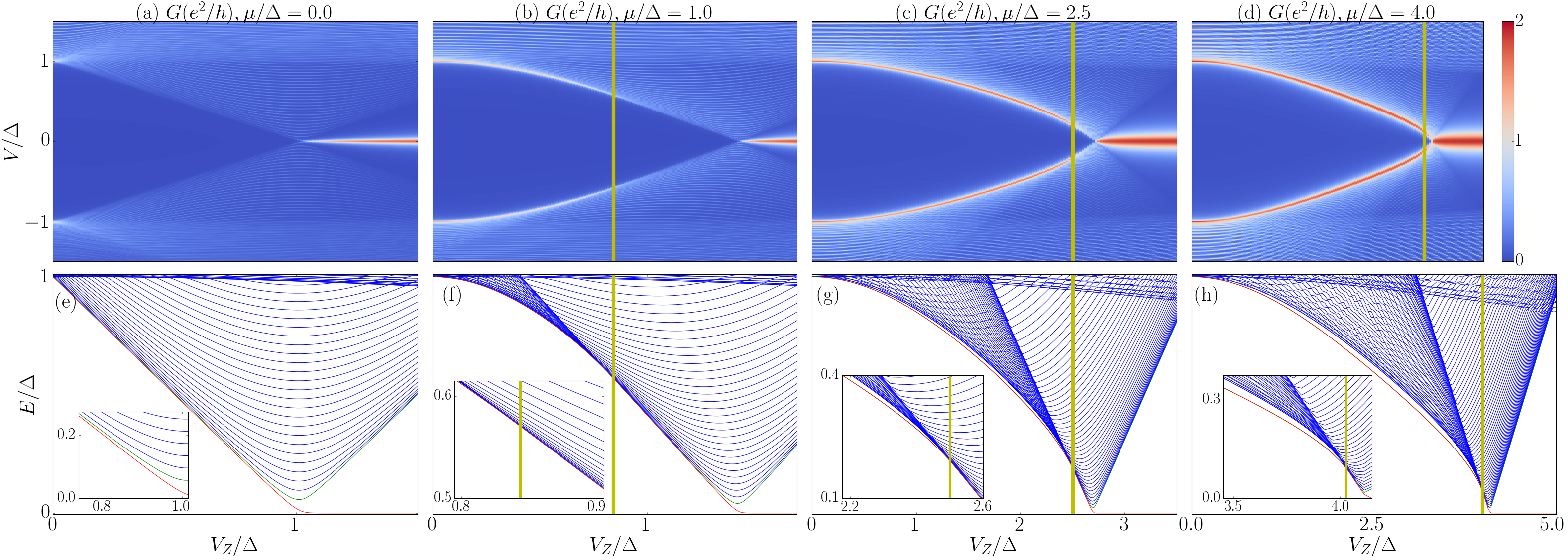}
\caption{(Color online) Upper panels: numerically calculated differential tunneling conductance through the Majorana nanowires as a function of Zeeman field for different chemical potentials. In (a), $\mu=0$, both ``gap closure" feature below TQPT and ``gap reopening" feature above TQPT are absent. In (b)--(d), at finite $\mu$,  a ``gap closure" feature at low Zeeman field becomes apparent below TQPT indicating the existence of intrinsic ABS, although the ``gap reopening" feature remains absent.
The termination of the apparent gap closing feature at $V_{Zt}$ ($V_{Zt}<V_{Zc}$, where $V_{Zc}=\sqrt{\mu^2+\Delta^2}$ is the TQPT) manifests the coming together of the intrinsic ABS. The location of $V_{Zt}$ is marked by a yellow vertical line in (b)--(d), which is analytically given by Eq.~\eqref{Eq:Vzt} in Sec.~\ref{sec:Analytical}B. Lower panels: the corresponding low-energy spectra as a function of Zeeman field for different chemical potentials. The red (green) lines correspond to the first (second) lowest energy eigenstates. When these two states are ABS localized at the opposite ends of the wire (i.e. for $V_Z < V_{Zt}$), they are degenerate as shown in (f)--(h). The blue lines denote the bulk states. Inset: zoom-in plot of energy spectra in the vicinity of $V_{Zt}$. The intrinsic ABS levels can be seen as the split-off states in the SC gap at finite $V_Z$ and finite $\mu$ slightly away from the bulk states. All the results are for nanowires length $L\cong20\mu$m, SC gap $\Delta=0.2$meV, and spin-orbit coupling $\alpha_R=0.5$eV$\angstrom$.}
\label{fig1}
\end{figure*}
The Bogoliubov--de Gennes (BdG) Hamiltonian for the one-dimensional superconductor-semiconductor hybrid nanowire is~\cite{Sau2010Generic, Lutchyn2010Majorana, Oreg2010Helical} $\hat{H}=\frac{1}{2}\int dx\hat{\Psi}^\dag(x)H_{NW}\hat{\Psi}(x)$,
\begin{equation}
H_{NW} = \left( -\frac{\hbar^2}{2m^*} \partial^2_x -i \alpha_R \partial_x \sigma_y - \mu \right)\tau_z + V_Z\sigma_x + \Delta_0 \tau_x,
\label{eq:H}
\end{equation}
where $\sigma_{\mu}(\tau_{\mu})$ are Pauli matrices in spin (particle-hole) space. Here, most parameters are chosen to be consistent with the experimental measurements~\cite{Lutchyn2017Realizing,Gul2018Ballistic,Kammhuber2016Conductance,Kammhuber2017Conductance}, e.g., the effective mass $m^*=0.015m_e$, the spin-orbit coupling $\alpha_R=0.5~$eV$\angstrom$, the proximity-induced superconducting gap $\Delta=0.2$~meV, and  the Zeeman spin splitting energy $V_Z=g\mu_BB/2$ with Bohr magnetron $\mu_B$ and Land\'e factor $g\simeq40$. Note that the pristine nanowire we consider is without any quantum dot or chemical potential variation. Thus, there is no issue of any accidental or extrinsic ABS in our pristine system as considered in Refs.~\cite{Brouwer2011Topological,Chiu2017Conductance,Liu2017Andreev,Moore2016Majorana,Prada2012Transport}; any ABS found in our work is intrinsic to the pristine nanowire and is not induced by extrinsic disorder (e.g., accidental quantum dots, chemical potential variations). In order to calculate the differential conductance ($G=dI/dV$) through the Majorana nanowire, we set up a normal-metal-superconductor (NS) junction, for which the normal-metal lead Hamiltonian contains only the first three terms in Eq.~\eqref{eq:H}, and the zero-bias voltage chemical potential of the lead is set as $\mu_\text{lead} = 25~$meV. In addition, a tunnel barrier of height $E_\text{barrier}=10~$meV and width $l_\text{barrier}=20~$nm lies at the junction interface~\cite{Liu2017Role}. The differential conductance is calculated by the $S$-matrix method~\cite{Blonder1982Transition, Setiawan2015Conductance}. The elements of $S$ matrix are computed using a Python package KWANT~\cite{kwant}, for which we first discretize the BdG Hamiltonian \eqref{eq:H} into a tight-binding form. Similarly, to get the energy spectra and the corresponding low-energy wave functions of the Majorana nanowire, we diagonalize the tight-binding version of the BdG Hamiltonian \eqref{eq:H}. We obtain results by varying $V_Z$, the Zeeman spin splitting, and $\mu$, the chemical potential, keeping other parameters ($\Delta$ and $\alpha$) fixed in the system, using a large value of the wire length $L\cong20\mu$m (except in Sec.~\ref{sec:shorter}, where we discuss experimental implications) so that the MBS are effectively well separated to avoid complications from any possible Majorana oscillations or discrete states arising from size quantization in the wire.  Our finding is thus valid in the thermodynamic limit of very long wires, and is not a finite-size artifact.

\section{calculated differential conductance and energy spectra}\label{sec:numerical}
We start by discussing the differential conductance ($G=dI/dV$) for the NS junction. The numerical simulations for the differential conductance as a function of the Zeeman field ($V_Z$) at various chemical potentials ($\mu$) are shown in the upper panels of Fig.~\ref{fig1}. Figure~\ref{fig1}(a) shows the differential conductance at zero chemical potential ($\mu=0$), which is the widely studied case in the literature. As the Zeeman field increases from zero, the superconducting gap shrinks monotonically from a finite value $\Delta=0.2$~meV to smaller values. When the Zeeman field goes through the TQPT at $V_{Zc} =\Delta$, the gap completely closes and then reopens as predicted theoretically~\cite{Sau2010Generic}. At the same time, a ZBCP with a quantized value $2e^2/h$ forms indicating the formation of MBS inside the hybrid nanowire. Note that the gap closure pattern below TQPT and the gap reopening pattern above TQPT are both extremely weak features in the tunneling conductance spectra, but the MBS feature is prominent as a strong ZBCP~\cite{Stanescu2012To}. Such apparent invisibility of the differential conductance associated with gap closing/opening features is because the corresponding nanowire SC states are extended bulk states inside the quasiparticle continuum. Hence, as pointed out already~\cite{Stanescu2012To}, these bulk states do not couple well to the tunneling lead which mostly probes localized bound states at the wire end, leading to prominent MBS features as the ZBCP above the TQPT, but only very weak features corresponding to gap closing/opening below/above the TQPT. This is an unavoidable tunneling matrix element effect making only the localized end states in the nanowire to be particularly sensitive in the conductance spectrum. However, in apparent disagreement with this tunneling matrix element arguments, Figs.~\ref{fig1}(b)--\ref{fig1}(d) showing the differential conductance through Majorana nanowires with finite chemical potentials ($\mu\neq0$) manifest strong gap closing features, but no gap opening features, in the conductance as $V_Z$ sweeps through the TQPT. The stark difference between the results in Fig.~\ref{fig1}(b)--\ref{fig1}(d) for finite $\mu$ and that in Fig.~\ref{fig1}(a) for the zero chemical potential case is a strong gap closing pattern denoted by a coherence peak below TQPT, although above TQPT the gap reopening feature is still very weak in both situations. Such a strongly asymmetric visibility in conductance between gap closing and reopening patterns for the differential conductance is consistent with multiple recent experimental observations~\cite{Zhang2017Quantized, Deng2016Majorana,Nichele2017Scaling} where typically the gap closing (opening) conductance feature is strong (weak).

Actually, the coherence peak associated with the gap closure manifesting in Figs.~\ref{fig1}(b)--\ref{fig1}(d) arises from the presence of \emph{intrinsic} localized ABS in the topologically trivial regime, which exist for any nonzero $\mu$ and $V_Z$; these intrinsic ABS disappear for $\mu=0$, or  $V_Z=0$, or $V_Z>V_{Zt}$. These intrinsic ABS are the subject matter of this study. We emphasize that this qualitative difference in the field-dependent conductance behavior between zero $\mu$ [Fig.~\ref{fig1}(a)] and nonzero $\mu$ [Figs.~\ref{fig1}(b)--\ref{fig1}(d)] happens only in the trivial regime ($V_Z<V_{Zc}$) below the TQPT; above the TQPT ($V_Z>V_{Zc}$), the two situations (zero and nonzero chemical potentials) are qualitatively similar with the ABS transmuting into MBS as a strong ZBCP with large conductance and the topological gap opening being a weak feature.

We first study how the gap closure pattern evolves with increasing chemical potential. Figures~\ref{fig1}(b)--\ref{fig1}(d) show the differential conductance for $\mu$ increasing from $\Delta$ to $4\Delta$. For low values of chemical potential ($\mu=\Delta$), as shown in Fig.~\ref{fig1}(b), the coherence peak at the edge of quasiparticle continuum is much smaller ($\ll 2e^2/h$) than that of MBS-induced ZBCP although it is still stronger than the corresponding situation for zero $\mu$. We note that for small $\mu$, the range of $V_Z$ over which such a strong coherence peak exists is rather small, i.e., the coherence peak disappears well below the TQPT for small $\mu$.
We call the value of the Zeeman field $V_{Zt}$, at which the ABS coherence peak vanishes [marked by the yellow vertical lines in Figs.~\ref{fig1}(b)--\ref{fig1}(d)]. This is a characteristic Zeeman field which has not been reported before in the literature. The characteristic Zeeman field $V_{Zt}$ defines the regime of existence ($0<V_Z<V_{Zt}<V_{Zc}$) of the intrinsic ABS giving rise to the coherence peak below the TQPT as discussed below. In fact, the coherence peak for $V_Z<V_{Zt}$ arises specifically from the intrinsic ABS, which always exist provided $\mu$ is finite, but for small $\mu$, $V_{Zt}$ tends to be small also and may not be prominent in the conductance spectrum. By contrast, for a large chemical potential ($\mu=2.5\Delta$ or $4\Delta$), as shown in Figs.~\ref{fig1}(c) and \ref{fig1}(d), the ABS-induced conductance peak can be as high as the MBS-induced ZBCP, i.e., approaching a conductance value of $2e^2/h$. Increasing $\mu$ leads to $V_{Zt}$ approaching $V_{Zc}$, thus making it difficult to distinguish between $V_{Zt}$ and $V_{Zc}$ at large $\mu$, particularly in a realistic situation involving finite-energy resolution arising from temperature, disorder~\cite{Liu2012Zero, Mi2014X, Sau2013Density}, dissipation, and level broadening. Thus, for generic $\mu$ (which is by definition large compared to the induced gap) $V_{Zt}$ may be very close to $V_{Zc}$, and the strong peak in the conductance arising from the intrinsic ABS may dominate the apparent gap closing conductance feature on the trivial side throughout the $V_Z<V_{Zc}$ regime. In fact, this is precisely the experimental observation in the best available data~\cite{Zhang2017Quantized, Deng2016Majorana,Nichele2017Scaling}, where the strong conductance feature associated with the gap closing on the low-field side (i.e., $V_Z<V_{Zc}$, where the gap closes and the ZBCP shows up) is always present in a prominent manner in conflict with the theoretical prediction at zero $\mu$ which predicts that no gap closing feature should manifest itself.

To further characterize the ABS, we now present the corresponding energy spectra of the Majorana nanowire in the lower panels of Fig.~\ref{fig1}. In the zero chemical potential case [Fig.~\ref{fig1}(e)], the lowest excitation energy in the topologically trivial regime ($V_Z < V_{Zc}$) belongs to the quasi-particle continuum (blue lines), and therefore the corresponding energy eigenstates are extended bulk states. This has been the belief in much of the existing Majorana literature, i.e., the gap closing states on the trivial side ($V_Z<V_{Zc}$) are extended bulk continuum states which come together with increasing Zeeman field at the TQPT ($V_Z=V_{Zc}$) leading to the vanishing of the bulk nontopological gap. On the other hand, when the chemical potential is finite [as shown in Figs.~\ref{fig1}(f)--\ref{fig1}(h)], there are two doubly degenerate low-energy modes (in red and green colors) below the quasi-continuum and separated from the bulk continuum by a small but finite energy gap. Thus, there exists a pair of in-gap localized ABS in the trivial regime for finite chemical potential nanowires, which are responsible for the conductance peaks observed in Figs.~\ref{fig1}(b)--\ref{fig1}(d). In addition, at the characteristic Zeeman field $V_{Zt}$, the two bound-state energies merge into the quasiparticle continuum, indicating that the two ABS come together and become bulk states at $V_{Zt}$ below the TQPT (but arbitrarily close to $V_{Zc}$ for large chemical potential values). Thus, beyond $V_{Zt}$, this conductance peak vanishes, as bulk states do not contribute to the differential conductance due to the small tunneling matrix elements coupling bulk states with the tunneling leads at the ends of the wire. We note that the ABS disappear completely on the topological side (Fig.~\ref{fig1}) where the only states are the zero-energy midgap MBS and finite-energy bulk continuum states; thus, one can think of the intrinsic ABS transmuting to topological MBS since the extended bulk continuum states exist on both sides of $V_{Zc}$ (and are not the lowest-energy states on either side except at zero chemical potential). We mention that these ABS are somewhat reminiscent of exciton states in semiconductors, where the exciton, a bound electron-hole state in contrast to the extended bulk Bloch band states, also exists inside the semiconductor band gap (whereas the ABS here exists inside the bulk SC gap) and carries substantial spectral weight for band-to-band optical transitions.  One can loosely think of these intrinsic ABS in the nanowire as being similar to intrinsic exciton states in semiconductors.

\section{wave function and Phase diagram}\label{sec:phase}
\begin{figure}
\centering
\includegraphics[width=1.0\columnwidth]{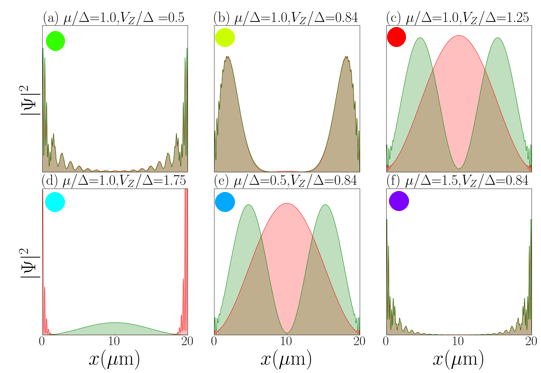}
\caption{(Color online) The spatial distribution of the wave functions of the first (second) lowest-energy eigenstates in red (green) colors at various chemical potentials and Zeeman fields. Panels (a)--(d) are at fixed $\mu/\Delta=1$ with different $V_Z$, while panels (b), (e), and(f) are at fixed $V_Z/\Delta=0.84$ with different $\mu$. In particular, states in (a) and (f) are all ABS, states in (c) and (e) are all extended bulk states. (b) $V_Z=V_{Zt}$. (d) $V_Z>V_{Zc}$, the two lowest states are one MBS and a bulk state. The colored dots of each panel indicate their position in $(\mu/\Delta,V_Z/\Delta)$ plane in Fig.~\ref{fig3}(a). The other parameters are the same with Fig.~\ref{fig1}.}
\label{fig2}
\end{figure}

In this section we numerically investigate the wave functions of the first two lowest-energy eigenstates focusing on the localized ABS, the extended bulk states, and the transition among them (i.e., their coming together at $V_{Zt}$). We also analyze the first three eigenenergies of the Majorana nanowire and obtain the enriched phase diagram of the hybrid nanowire in light of our discovery of the intrinsic ABS in the trivial part of the phase diagram for $V_Z<V_{Zt}$. These energy-spectra analyses are complementary to the conductance calculation in the previous section, as the strong conductance peaks in the gap closing arise from localized ABS, while the weak conductance features in gap closing arise from extended bulk states. We note the obvious:  The topological side of the phase diagram is the same for zero and nonzero chemical potentials (i.e., only zero-energy midgap localized MBS and finite-energy extended bulk states above the SC gap exist), as described in the existing literature, and is not affected by the presence of intrinsic ABS on the nontopological side for finite $\mu$.

Figure \ref{fig2} shows the calculated wave functions for the two lowest-energy eigenstates in the nanowire at various parameter regimes. Specifically, Figs.~\ref{fig2}(a)--\ref{fig2}(d) show how these two states evolve with increasing Zeeman field at fixed chemical potential $\mu=\Delta=0.2$~meV. In Fig.~\ref{fig2}(a), when $0 < V_Z < V_{Zt}$, there are two well-localized ABS at each wire end which lead to large conductance peaks in the tunneling conductance. As the Zeeman field increases to $V_Z = V_{Zt}$, as shown in Fig.~\ref{fig2}(b), the envelopes of the two ABS at each wire end begin to leak into the middle region of the nanowire, and the ABS localization length becomes comparable to the nanowire size, thus becoming extended bulk states. When the Zeeman field goes through $V_{Zt}$ but still in the topologically trivial regime, i.e., $V_{Zt} < V_Z < V_{Zc}$, the lowest two energy eigenstates are both extended bulk states as shown in Fig.~\ref{fig2}(c). Finally, as the Zeeman field goes beyond the critical Zeeman field $V_Z > V_{Zc}$, the nanowire enters the topological regime and a pair of localized MBS form and the first excited state is a bulk state, as shown in Fig.~\ref{fig2}(d). Such a wave-function evolution is a precise microscopic manifestation of all the essential features in Figs.~\ref{fig1}(b)--\ref{fig1}(d), e.g., conductance peaks for gap closing ($V_Z < V_{Zt}$) arising from ABS, disappearance of conductance peaks ($V_Z = V_{Zt}$), weak gap closing ($V_{Zt} <V_Z < V_{Zc}$) due to extended bulk states, weak `gap reopening' ($V_Z >V_{Zc}$) due to bulk states on the topological side, as well as the MBS showing up as the ZBCP. We also show the evolution of the wave function as a function of chemical potential $\mu$ at fixed Zeeman field in Figs.~\ref{fig2}(b), \ref{fig2}(e), and \ref{fig2}(f). They show that at low (high) chemical potential, the two lowest-energy eigenstates are extended bulk states (localized ABS). We note that the intrinsic ABS exist for all finite $\mu$, except that their regime of existence in the magnetic field (i.e., the value of $V_{Zt}$) is suppressed as $\mu$ decreases, with the range of $V_Z$ for their existence eventually vanishing for vanishing $\mu$.

\begin{figure}
\centering
\includegraphics[width=1.0\columnwidth]{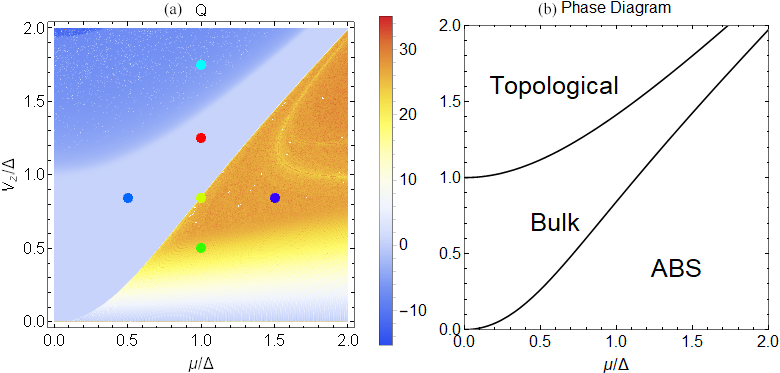}
\caption{(Color online) (a) The contour plot of $Q=\ln{\frac{E_3-E_2}{E_2-E_1}}$ as a function of $\mu$ and $V_Z$. The six colored dots mark the parameters of the wave functions shown in Fig.~\ref{fig2}.
(b) The ``phase diagram" showing the topological phase and the two distinct trivial regimes, the ABS regime and the extended bulk regime. The crossover between the ABS and the bulk regimes is given by Eq.~\eqref{Eq:Vzt} in Sec.~\ref{sec:Analytical}B, while the boundary between the bulk state regime and the topological regime satisfies $V_{Zc}=\sqrt{\Delta^2+\mu^2} $.}
\label{fig3}
\end{figure}
The careful investigation of the wave functions of the two lowest-energy eigenstates motivates us to define a ``phase diagram" based on the low-energy wave-function features. The phase diagram contains three distinct regions: First, the topological phase at high field ($V_Z>V_{Zc}$) where the first single-particle eigenstate is the localized MBS and the second and third states are both extended bulk states with a finite-energy gap (i.e., the SC gap on the topological side) above the MBS. Second, the trivial bulk regime, where the first three states are all extended bulk states above the SC gap. Third, which is the finding in this paper, the \emph{intrinsic} ABS regime, where there are two degenerate ABS localized at wire ends, and the third state is an extended bulk state separated from the two ABS by a finite-energy gap. To further characterize the three distinct regions in a quantitative way, we propose a quantity
\begin{align}
Q=\ln{\frac{E_3-E_2}{E_2-E_1}},
\end{align}
where $E_i~(i=1,2,3)$ is the $i$--th lowest eigenenergy for the Majorana nanowire. Note that the scaling behavior of energy difference as a function of wirelength $L$ is $\Delta E \sim e^{-L}$ for two doubly degenerate localized ABS, $\Delta E \sim L^{-1}$ for two extended bulk states, and $\Delta E \sim \text{const}$ for states separated by finite energy gap. Therefore, the distinction among the three regions can be summarized as
\begin{equation}
Q \sim
\begin{cases}
 \ln{\frac{L^{-1}}{E_\text{gap}}}<0& \text{ topological region},\\
 \ln{\frac{L^{-1}}{L^{-1}}} \sim 0& \text{ extended bulk state region}, \\
 \ln{\frac{L^{-1}}{e^{-L}}} \gg0& \text{ ABS region, for extremely long wire}.
\end{cases}
\end{equation}
The corresponding value of $Q$ in the phase diagram of chemical potential and Zeeman field is shown in Fig.~\ref{fig3}(a). We can clearly see the three distinct regimes arising from the numerically calculated $Q$ values: ABS (orange), bulk (light blue), topological (dark blue). Note that at zero chemical potential, there is no \emph{intrinsic} ABS (and hence no orange region), while at finite chemical potential the system generically has three ``phases", i.e., ABS, bulk, and topological ``phases". When the Zeeman field goes to even larger values, the extended bulk state region (i.e., light blue) shrinks, which is consistent with the calculated differential conductance in Fig.~\ref{fig1}. Also note that at zero Zeeman field, there is no ABS (we will give proof in Sec.~\ref{sec:Analytical}). At finite but small Zeeman field, the region in the phase diagram is still blue instead of orange. This is because the energy separation between the ABS and the bulk states scales to zero when Zeeman field goes to zero. Thus, although the horizontal line of $V_Z=0$ is the bulk ``phase", the region of $V_Z>0$ and $\mu \neq 0$ is in principle ABS ``phase" (more clarification in the Sec.~\ref{sec:Analytical}). Asymptotically, therefore, there are only two ``phases" for small and large $\mu$: only bulk and topological for small $\mu$ (this is already well known) and only ABS and topological for large $\mu$. For a generic system ($\mu=0$ is non-generic), there are three ``phases", but since high $\mu$ is the generic situation, the light blue region typically has a very small range of allowed $V_Z$ values since $V_{Zt}\sim V_{Zc}$ generically.

The ``phase diagram" of Fig.~\ref{fig3} is the qualitative result of our work, which we now study analytically below. We emphasize that our finding remains valid in the thermodynamic limit, and increasing $L$ further does not change our results (as we have explicitly verified numerically).

\section{Analytical understanding of the intrinsic Andreev bound states }\label{sec:Analytical}

This section is devoted to understanding the mechanism for the existence of Andreev bound states from the analytical perspective. We will show that intrinsic ABS exist in the topologically trivial regime as long as the chemical potential is finite. We further give the analytical expression for $V_{Zt}$, i.e., the range of the existence of ABS. These analytical results are in perfect agreement with the numerical results presented in Sec.~\ref{sec:phase} above.

\subsection{Existence of intrinsic Andreev bound states }

Here, we demonstrate that for a semi-infinite-long nanowire ($x>0$), there is always a bound state in the topologically trivial regime for finite $V_Z$ and $\mu$. The Hamiltonian we use is identical to Eq.~\eqref{eq:H}, which is
\begin{equation}\label{BdG}
H(k)=(k^2-\mu+k\sigma_z)\tau_z+V_Z\sigma_x+\Delta\tau_x,
\end{equation}
where $\frac{\hbar^2}{2m^*}= \alpha_R=1$ for notational simplicity, $k=-i\partial_x$~\cite{DasSarma2012Splitting}. Note that here the spin-orbit coupling is along the $z$ axis as we rotate the spin basis by $U = e^{-i\pi \sigma_x/4}$ without loss of generality. Since a general wave function is the superposition of all the particular solutions, we first seek eigenfunctions of the form $ \psi_n(x)=e^{-i k_n x}u_n $ with eigenenergy $ E $ satisfying
\begin{equation}\label{eig}
\left[ H(k_n)-E\right]u_n=0
\end{equation}
where $ u_n $ is a four-dimensional Nambu spinor. A symmetry analysis of the Hamiltonian shows that as the Hamiltonian is real and $ \sigma_xH(k)\sigma_x=H(-k) $, if  $ (k_n,u_n) $ is a solution, so is $ (k_n^*,u_n^*) $ and $ (-k_n,\sigma_x u_n) $. Since we are considering Andreev bound states in our problem, the normalizability condition $ \int_0^\infty dx |\psi(x)|^2<\infty $ constrains that among the eight solutions to Eq.~\eqref{eig}, we should only consider the solution in the lower half of the complex plane, namely $ k_1,-k_1^*,k_2,-k_2^* $ [$ \text{Im}(k_1)<0, \text{Im}(k_2)<0 $], along with their eigenvectors $ u_1,\sigma_x u_1^*,u_2,\sigma_x u_2^* $. The boundary condition we impose on the wave function in the semi-infinite wire is that $ \psi(0)=\sum_n C_nu_n=0 $, which means that the four eigenvectors $ u_1,\sigma_x u_1^*,u_2,\sigma_x u_2^* $ are linearly dependent. This motivates us to write a determinant whose value is zero:
\begin{equation}\label{F(E)}
F(E)=\det\left[u_1;u_2;\sigma_x u_1^*;\sigma_x u_2^*\right]=0.
\end{equation}
In addition, it is easy to show that $ F(E) $ is real, as

\begin{figure}
\centering
\includegraphics[width=\linewidth]{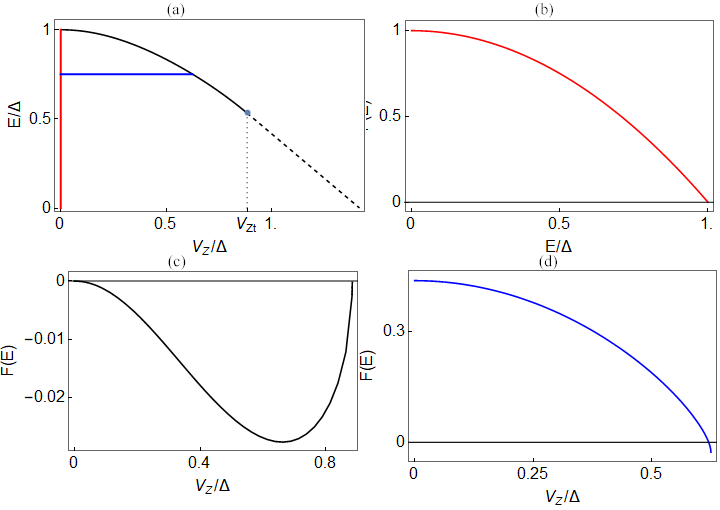}
\caption{(a) The bulk SC gap (black line) as a function of $V_Z$ at $\mu=\Delta$. Its solid part and dashed part are separated by the blue point at $V_Z=V_{Zt}$. Along the solid lines in different colors in (a), $F(E)$ are calculated in (b)--(d). (b) $F(E)$ as a function of $E$ along $V_Z=0$. For $E<\Delta$, $F(E)>0$. (c) $F(E)$ as a function of $V_Z$ along the SC gap below $V_{Zt}$ [the black solid line in (a)]. For $0<V_Z<V_{Zt}$, $F(E)<0$. (d) $F(E)$ as a function of $V_Z$ at fixed $E=0.75\Delta$. $F(E)$ is monotonic and crosses zero only once.  The other parameters are the same with Fig.~\ref{fig1}.
}
\label{fig4}
\end{figure}

\begin{align}\label{positive}
F(E)^*&=\det\left[u_1^*;u_2^*;\sigma_xu_1;\sigma_xu_2\right] \\ \nonumber
&=\det\left[\sigma_xu_1^*;\sigma_xu_2^*;u_1;u_2\right] \\ \nonumber
&=F(E).
\end{align}
Therefore, a crucial deduction from the previous analysis is that the bound-state solution occurs whenever $F(E)$ changes the sign. So, in the following, we will fix some $E$ inside the SC gap, and calculate the corresponding $F(E)$ horizontally from $V_Z = 0$ all the way to the SC gap [see blue line Fig.~\ref{fig4}(a)] to see whether $F(E)$ would flip the sign. Or an equivalent and more efficient way is to calculate $F(E)$ only at $V_Z = 0$ and at SC gap to see whether they show opposite sign. If opposite, there must be an ABS solution on the horizontal line connecting the two points. When $V_Z=0 $, using the symmetry $ \left[H,\sigma_z\right]=0 $, we can easily obtain $u_{1,2}=(r \pm i\sqrt{1-r^2},1,0,0)^T $ for an in-gap solution ($ r=E/\Delta<1 $). The corresponding $ F(E) $ is
\begin{equation}\label{key1}
F(E)=4\left(1-r^2\right)=\frac{4(\Delta^2-E^2)}{\Delta^2}>0,
\end{equation}
which indicates that there is no in-gap bound state for $ E<\Delta $ at zero Zeeman field. This is of course the known result at zero Zeeman energy where all solutions of the BdG equation are bulk extended above-gap continuum states. When the Zeeman field is turned on perturbatively, starting from the two eigenfunctions $u_1 =u_2=(1,1,0,0)^T$ at the SC gap $E=\Delta$, the perturbed wave function is $u'_1 = (1,1,a,b)^T$ and $u'_2 = (1,1,c,d)^T$ with real numbers $a,b,c,d$ of $O(V_Z)$. Thus, the $F(E)$ becomes
\begin{align}
F(E) = \det\left[\sigma_x u'_1;\sigma_x u'_2;u'^*_1;u'^*_2\right] = (a-b-c+d)^2 < 0,
\end{align}
up to $O(V^2_Z)$. Since $F(E)$ flips sign when $V_Z$ is turned on, we conclude that there must be a bound state just below the SC gap at small Zeeman field. So, we prove that a bound state, i.e., \emph{intrinsic} ABS, always forms for a class D nanowire whenever the Zeeman field is turned on. This is exactly what we found in our numerical results presented in Secs.~\ref{sec:model}--\ref{sec:phase}. Note that the presence of spin-orbit coupling, Zeeman splitting, finite chemical potential, and superconductivity are all necessary for producing the ABS in the nanowire in the trivial regime.

\begin{figure}
	\includegraphics[width=\linewidth]{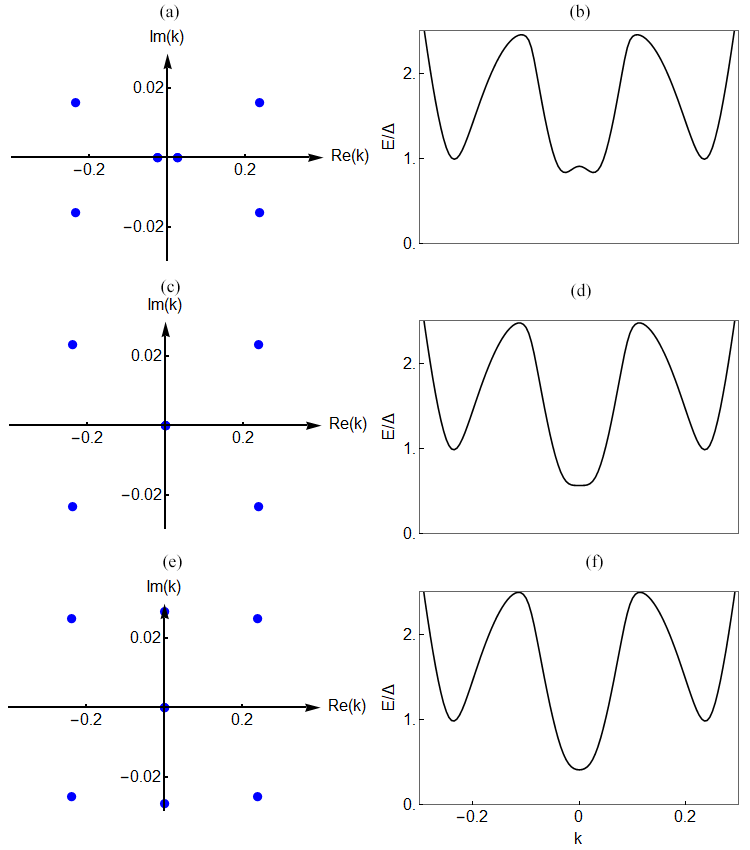}
	\caption{The complex $k_n$'s of the solutions to Eq.\eqref{eig} at the SC gap ($E=E_\text{gap}$) in the complex plane (left panels) and the lower positive energy band of the spectrum as a function of momenta (right panels) at $\mu/\Delta=1$ for different Zeeman fields. (a) At $V_Z=0.5\Delta<V_{Zt}$, the $k_n$'s at the two points on the real axis of the complex plane are doubly degenerate. The two $k_n$'s correspond to the two global band minima in (b). (c) At $V_Z=V_{Zt}$, the $k_n$'s at the origin point are quadruply degenerate. The four $k_n$'s correspond to the band minimum at $k=0$ in (d). (e) At $V_Z=\Delta>V_{Zt}$, the $k_n$'s at $k=0$ are doubly degenerate, while another two $k_n$'s are on the imaginary axis. (f) The corresponding minimum of the band is at $k=0$. The other parameters are the same with Fig.~\ref{fig1}.
}
\label{fig5}
\end{figure}

\subsection{Zeeman range of Andreev bound states}
In the previous subsection, using the perturbation theory we prove that an in-gap ABS always exists just below the SC gap for any infinitesimal Zeeman field. Now we address analytically the range of Zeeman field (i.e., $V_{Zt}$) over which the intrinsic ABS exist. When the Zeeman field is finite but not small, the perturbative argument breaks down, and therefore we need to numerically calculate $F(E)$. Figure~\ref{fig4} shows the bulk gap and the value of $F(E)$ along some specific path. In Fig.~\ref{fig4}(a), the black line denotes the bulk gap of the Majorana nanowire. Our logic is that if $F(E)$ at the bulk gap has opposite sign with respect to $F(E)$ at $V_Z=0$, there must exist some $V_Z$ in between giving $F(E)=0$ and thus leading to an in-gap ABS.  As a starting point, we calculate $F(E)$ along $0<E<\Delta$ at $V_Z=0$, which is the regime where we already have the analytic solution. As shown in Fig.~\ref{fig4}(b), the corresponding $F(E)$ is always positive along this line, which is consistent with our conclusion in Eq.~\eqref{key1}. In Fig.~\ref{fig4}(c), we calculate $F(E)$ at the SC gap (the solid part of the black line). With increasing $V_Z$, $F(E)$ is negative until $V_{Zt} \simeq 0.84 \Delta$, beyond which $F(E)$ becomes complex and not well defined. Thus, for $0 < V_Z < V_{Zt}$, as we argue, there always exists an in-gap ABS in the nanowire.

To get the analytic expression for $V_{Zt}$, we need to trace the eight roots of Eq.~\eqref{eig} in the complex plane. When $E$ is on the SC gap and $V_Z<V_{Zt}$ [on the black line in Fig.~\ref{fig4}(a)], there are two doubly degenerate $k_n$'s on the real axis [Fig.~\ref{fig5}(a)]. The two real $k$'s are associated with the two global band minima at finite momentum [Fig.~\ref{fig5}(b)], and the corresponding $F(E)$ is real and negative. When $V_Z$ increases, the two real $k$'s come closer to each other and they coincide at $k=0$ when $V_Z=V_{Zt}$ [blue point in Fig.~\ref{fig4}(a)]. Since $k=0$ is quadruply degenerate, $F(E)=0$. In the corresponding band structure, $k=0$ switches from local maximum to plateau. When $V_Z>V_{Zt}$, $k=0$ becomes doubly degenerate, and there are a pair of $k$'s on the imaginary axis, therefore, $F(E)$ is no longer guaranteed to be real. So,
 such a root analysis gives the condition for $V_{Zt}$, i.e., when the effective mass at $k=0$ is zero:
\begin{equation}\label{Eq:partial}
\eval{\frac{\partial^2 E}{\partial k^2}}_{k=0}=0.
\end{equation}
Or, equivalently, we can set the octic equation of Eq.~\eqref{eig} to be:
\begin{align}
a k^8+bk^6+ck^4+dk^2+e=0\quad \text{with}\quad d=0\quad \text{and}\quad e=0
\end{align}
such that the solution $k=0$ is quadrupolar degenerate. Both give the same analytic expression for $V_{Zt}$:
\begin{align}\label{Eq:Vzt}
V_{Zt}={V_{Zc}}\frac{-(\alpha_R^2 m^*-4\mu\hbar^2 )+ \sqrt{(\alpha_R^2m^*-4\mu \hbar^2)^2+\frac{16\alpha_R^2 \mu^3m^*\hbar^2}{V_{Zc}^2}}}{8\mu \hbar^2}.
\end{align}

 This characteristic Zeeman field $V_{Zt}$ is an important finding in our work, which leads to the classification of the three `phases' shown in Fig.~\ref{fig3}(b). When $V_Z<V_{Zt}$ there is \emph{intrinsic} ABS in the nanowire, while when $V_{Zt} < V_Z < V_{Zc}$ the low-energy states are all extended bulk states. When $V_Z>V_{Zt}$, the phase is the topological superconductivity with MBS. Finally we discuss the limiting behavior of $V_{Zt}$. When $\mu\rightarrow0$, $V_{Zt}\rightarrow0$, meaning there is no ABS for $\mu=0$. When $\mu \rightarrow \infty$, $V_{Zt} \rightarrow V_{Zc}$, meaning the bulk region between the ABS and topological phase vanishes asymptotically. For finite $\mu$, $V_{Zt} < V_{Zc}$, but since generically $\mu$ should be arbitrarily large (unless one can tune it to a small value $\mu\ll\Delta$), we expect $V_{Zt}\sim V_{Zc}$.

The numerical (Sec.~\ref{sec:phase}) and the analytical (Sec.~\ref{sec:Analytical}) work together decisively to establish that not only are there Andreev bound states in the trivial regime for finite $\mu$, but also that these ABS are doubly degenerate states localized at the two ends of the wire, i.e., there are four individual MBS (two overlapping MBS forming an ABS at each end of the wire).  It is clear that these doubly degenerate ABS transmute into two MBS (localized at two ends) and two gap opening extended bulk states at the TQPT with all the conductance spectral weight being contained in the MBS beyond the TQPT.

\section{comparison with experiments and discussion}\label{sec:shorter}

\begin{figure}
\centering
\includegraphics[width=1.0\columnwidth]{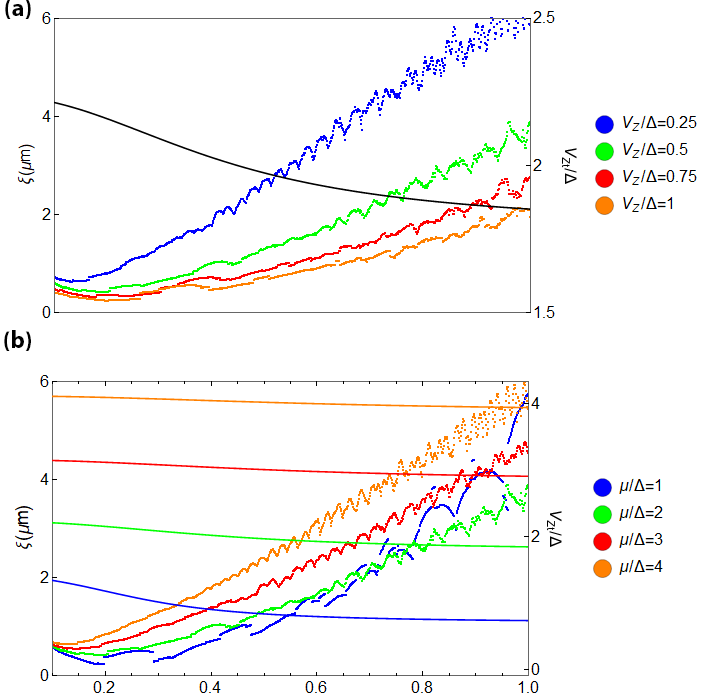}
\caption{\label{fig6}The ABS localization length $ \xi $ ($ \mu $m) in coarse dotted lines as the function of spin-orbit coupling strength $ \alpha $ (eV\AA). The wire is 20 $ \mu $m and superconducting gap $ \Delta=0.2 $meV. Smooth solid lines are $ V_{Zt} $, and ABS exists for $ V_{Z}<V_{Zt} $ (a) localization length plotted at a fixed chemical potential $ \mu/\Delta=2 $ for different $ V_Z/\Delta=0.25, 0.5, 0.75, 1 $; (b) localization length plotted at a fixed Zeeman splitting field $ V_Z/\Delta=0.5 $ for different $ \mu/\Delta=1, 2,3,4 $.}
\end{figure}
\begin{figure*}
\centering
\includegraphics[width=\textwidth]{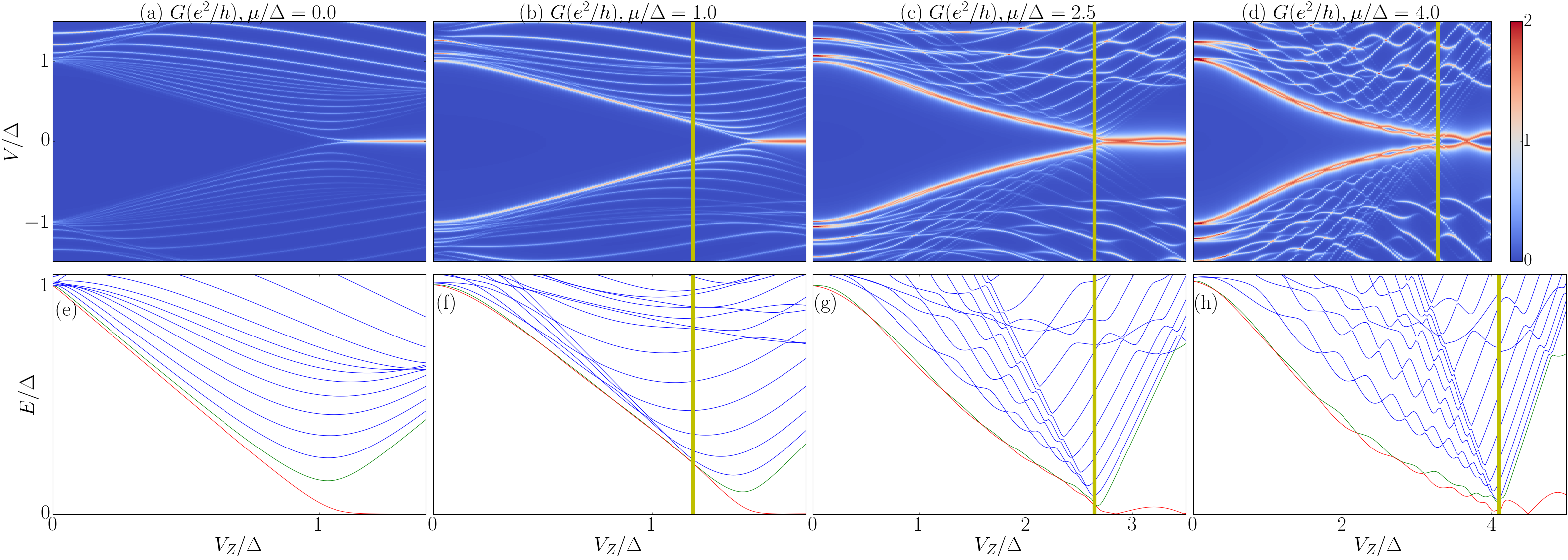}
\caption{\label{fig7}(Color online) Upper panels: numerically calculated differential conductance through short Majorana nanowires with length $L\cong3\mu$m, spin-orbit coupling $\alpha_R=0.2$eV$\angstrom$ and SC gap $\Delta=0.2$meV as a function of Zeeman field for different chemical potentials. The yellow vertical line marks the $V_{Zt}$ calculated by Eq.\eqref{Eq:Vzt} in infinite-long wire limit.
Note that the `gap closure' feature is strong in the short nanowires.
Lower panels: the corresponding low-energy spectra as a function of Zeeman field for different chemical potentials. While the blue lines denote the bulk states, the red and green lines are associated with the first and second lowest energy eigenstates respectively.
}
\end{figure*}
\begin{figure}
\centering
\includegraphics[width=1.0\columnwidth]{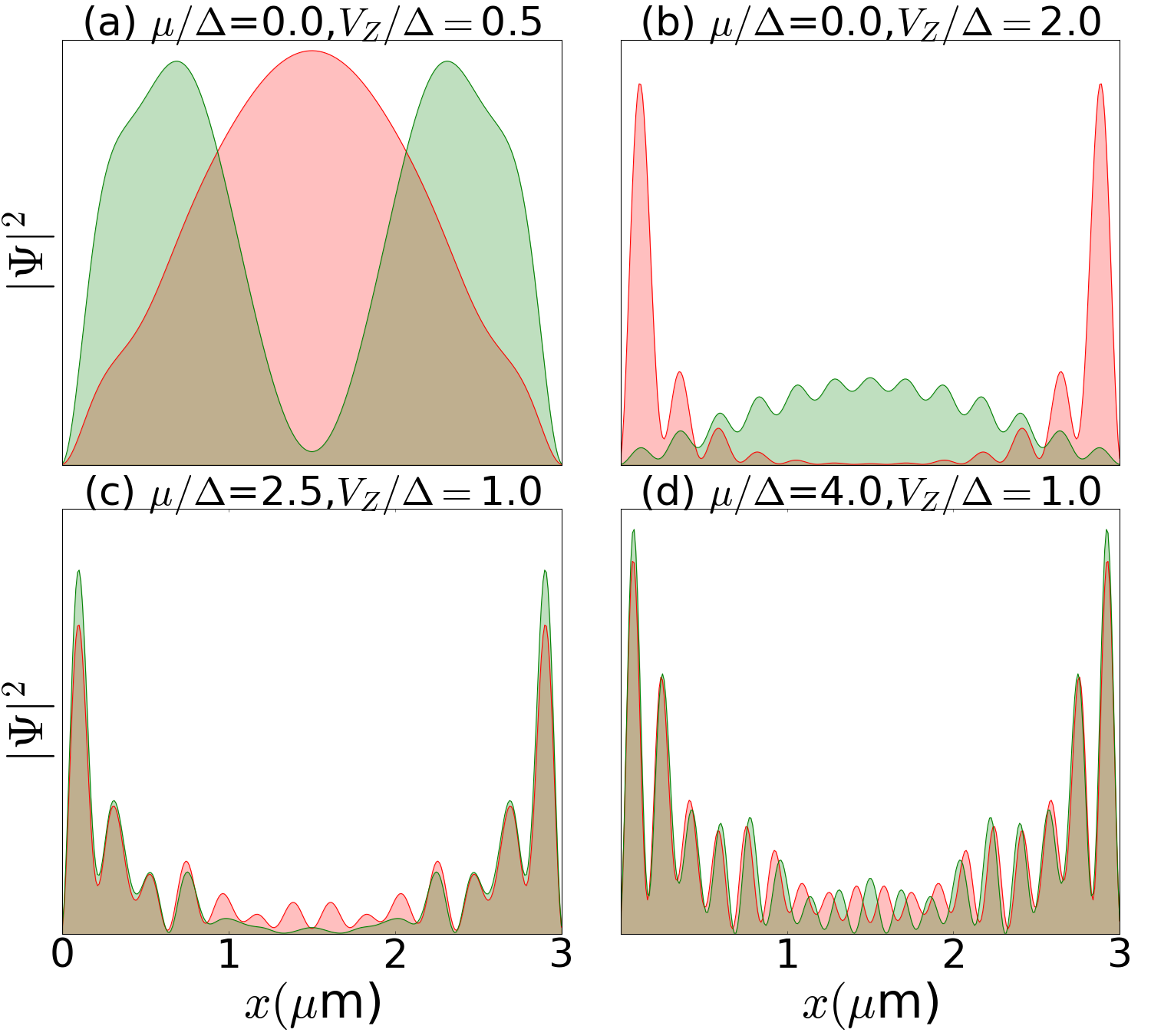}
\caption{\label{fig8}(Color online) The spatial distribution of the low-energy state wave functions in Fig.~\ref{fig7} at various chemical potentials and Zeeman fields. Upper panels: For $\mu$=0, (a) extended bulk states at $V_Z=0.5\Delta$($<V_{Zc}$) and (b) one MBS and one bulk state at $V_Z=2\Delta$($>V_{Zc}$). Lower panels:  ABS at $V_Z=\Delta$($<V_{Zt}$) for (c) $\mu=2.5\Delta$ and (d) $\mu=4\Delta$.
}
\end{figure}

Our discussion, both theoretical and numerical, has focused so far on the semi-infinite system (or a wire with $L\cong20\mu m$ for numerical simulations) corresponding to the long wire limit, so that finite-size artifacts are avoided, establishing clearly that the three `phases' we are discussing are all bulk `phases' and do not somehow arise from any finite-size quantization.  Thus, the doubly degenerate ABS for finite $\mu$ and $V_Z<V_{Zt}<V_{Zc}$ exist always independent of how long the nanowire might be. But, obviously, the localization length of the ABS must be smaller than the wire length $L$ for there to be any meaningful practical distinction between ABS and extended bulk states, even if conceptually they are very different (e.g., excitons versus band states in semiconductors). Equivalently, for short wires, the size quantization along the wire leads to all low-lying bulk states being size quantized bound states within the gap, and therefore, the practical distinction between the extended bulk states and localized ABS disappears.  We therefore discuss in this section the issue of the ABS localization length and the associated tunneling conductance spectra in shorter wires to make a comparison with the existing experimental literature.  Also, the corresponding energy spectra and wave functions are shown to characterize the ABS in short nanowire.

Obviously, the ABS localization length (i.e., the ABS wave function) depends on all the microscopic parameters entering the minimal BdG model Hamiltonian [Eq.~\eqref{eq:H}]: $\Delta$ (SC gap), $\alpha_R$ (spin-orbit coupling strength), $V_Z$ (Zeeman splitting), and $\mu$ (chemical potential). It turns out that the spin-orbit coupling is a key parameter determining the ABS localization length.  We depict in Fig.~\ref{fig6} our calculated ABS localization length as a function of the spin-orbit coupling strength for a few fixed values of $\mu$ [Fig.~\ref{fig6}(a)] and $V_Z$ [Fig.~\ref{fig6}(b)].  With the assumption that the outer envelope of an ABS wave function is an exponentially decaying curve, the ABS localization length at fixed $\mu$ and $V_Z$ can be obtained from the decay rate. To ensure the wave function at the ABS region (i.e., $V_Z<V_{Zt}$) for different values of $\alpha_R$, we plot the corresponding $V_{Zt}$ for reference. We find that in the wide range of chosen Zeeman fields and chemical potentials, when $\alpha_R$ increases from $0.2$eV\AA~to $1$eV$\angstrom$, the ABS localization length keeps increasing.  Experimentally, the precise $\alpha_R$ is not known, and it is likely that the applicable $\alpha_R$ varies from one sample to another since the Rashba coupling should depend on all the details of the spatial asymmetry in specific semiconductor-superconductor hybrid structures~\cite{Lutchyn2017Realizing}. It is clear, however, from Fig.~\ref{fig6} that the typical ABS localization length is comparable ($\sim$ few microns or less) to experimental wire lengths except for small values of chemical potential and Zeeman splitting.  We therefore expect most experimental samples to be in the ABS `phase', manifesting strong gap closing conductance features below the TQPT.

In Figs.~\ref{fig7} and \ref{fig8} we present results for spin-orbit coupling strength $\alpha_R=0.2$eV\AA~leading to rather small ABS localization length so that the effective ABS gap closing conductance features at finite chemical potentials are very strong [Fig.~\eqref{fig7}] with the corresponding energy spectra and wave functions shown in Figs.~\ref{fig7} and \ref{fig8} respectively.  Similar to the long wire case (upper panels of Fig.~\ref{fig1}), strong coherence peak appears in the trivial region at finite $\mu$ and manifests the `gap closure' feature in the short wire. Aside from, however, we can see finite size effect for short wire as well. As shown in Figs.~\ref{fig7}(c) and \ref{fig7}(d), the conductance peak below $V_{Zt}$ splits and oscillates. Correspondingly, in the lower panels of Fig.~\ref{fig7}, the two nearly degenerate lowest eigenenergies split and oscillate around each other below $V_{Zt}$. Such an oscillation in conductance and energy spectrum is due to the wave-function overlap between two ABS at wire ends for a finite-length nanowire. Figure~\ref{fig8}(c) and \ref{fig8}(d) show some typical localized ABS overlapping with each other in the middle region.

 The finite chemical potential results in Figs.~\ref{fig7} and\ref{fig8} may be directly compared qualitatively with the experimental data from  the groups in Delft ~\cite{Zhang2017Quantized} and Copenhagen~\cite{Deng2016Majorana} with both our theory and the experimental data manifesting strong gap closing features and strong ZBCP features without any discernible gap opening conductance features.  It is reasonable to conclude that the nanowire experiments are dominated by the ABS `phase' in the trivial regime before the ABS transmute to MBS at the TQPT.
Additional numerical results for representative system parameters are provided in Appendix A.

\section{Nanowires in the presence of external quantum dots}\label{sec:QD}

\begin{figure}
\raggedleft
\includegraphics[width=1.0\linewidth]{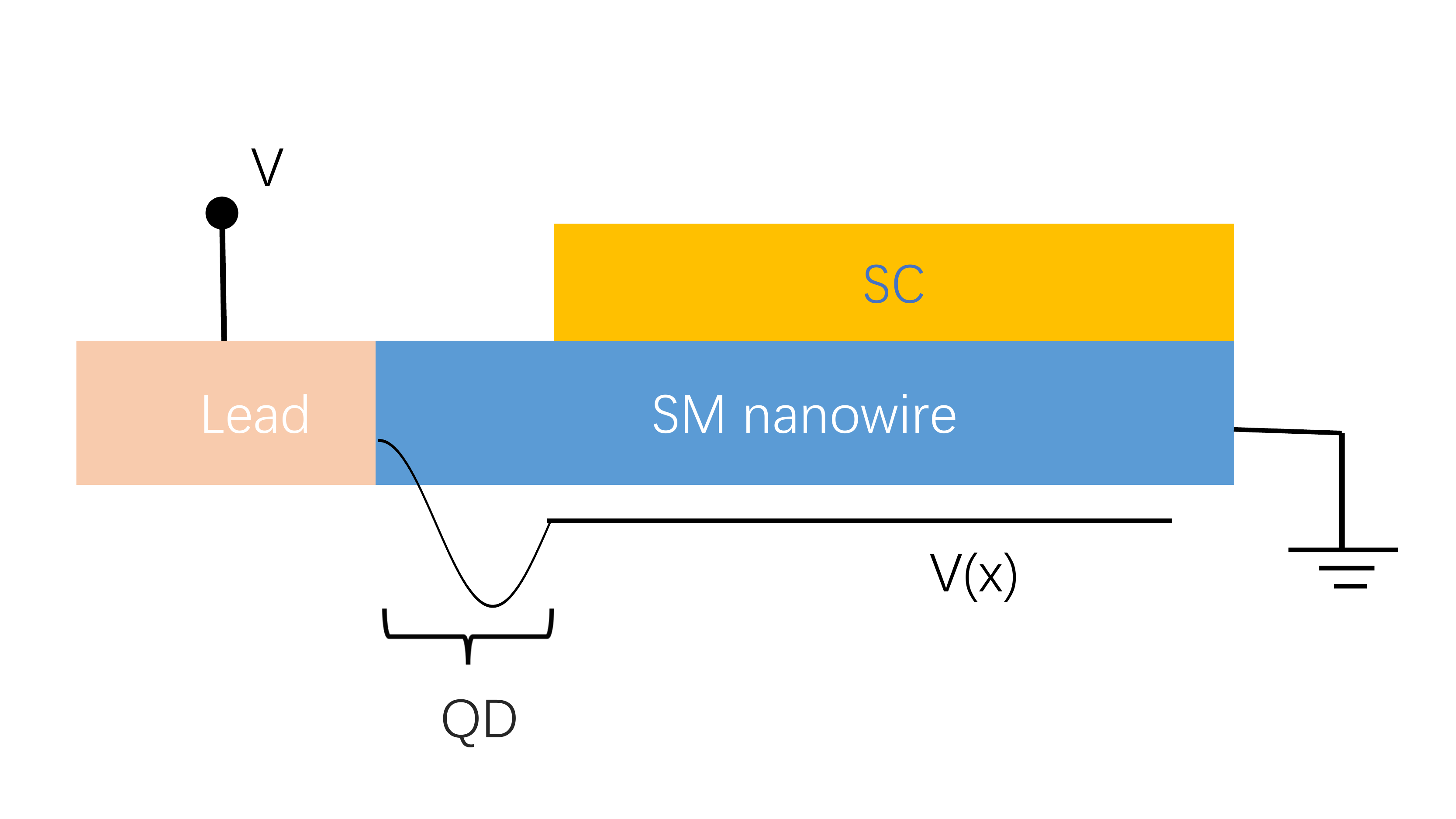}
\caption{(Color online) Schematic for NS junction with a non-SC quantum dot located between the lead and the nanowire. }
\label{figschQD}
\end{figure}

\begin{figure*}
\raggedleft
\includegraphics[width=1.0\textwidth]{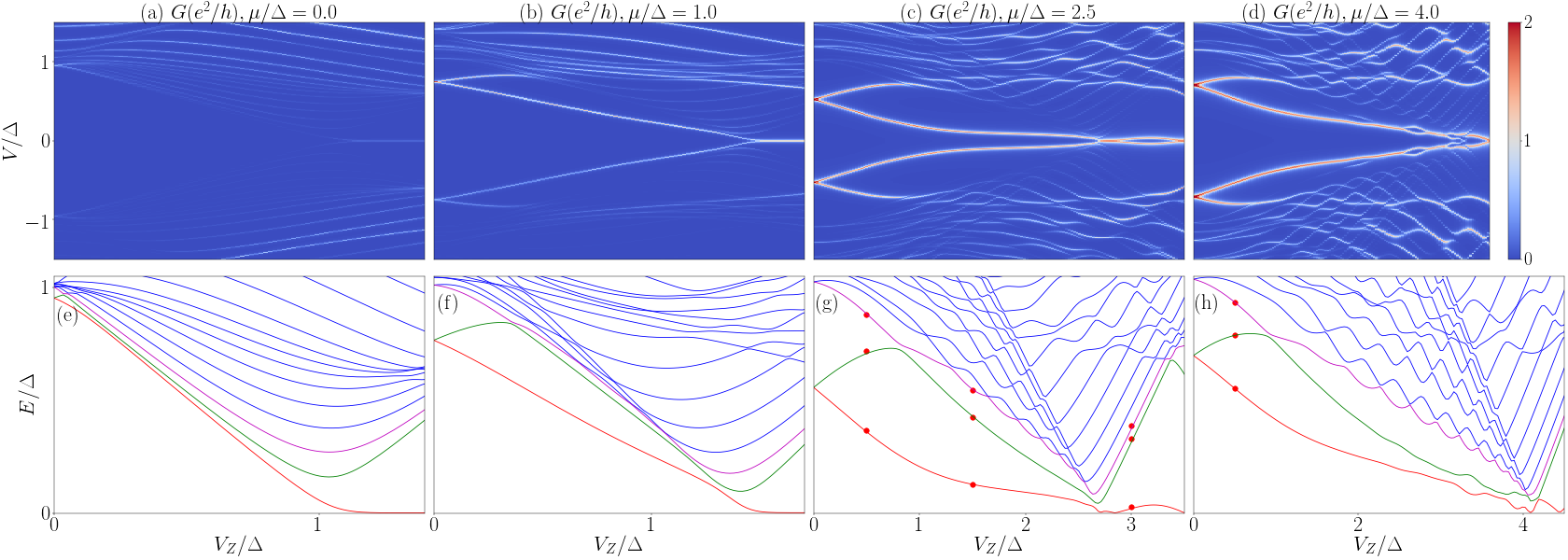}
\caption{(Color online) Upper panels: numerically calculated differential tunneling conductance through the dot-nanowire hybrid structure as a function of Zeeman field for different chemical potentials.  Lower panels: the corresponding low-energy spectra as a function of Zeeman field for different chemical potentials. The red, green, and purple lines correspond to the three lowes-energy eigenstates, respectively, while the blue lines denote the bulk states.  The red dots mark the eigenstates of which the wave functions are shown in Fig.~\ref{fig10}. All the results are for nanowires length $L\cong3\mu$m, SC gap $\Delta=0.2$meV, and spin-orbit coupling $\alpha_R=0.2$eV$\angstrom$.}
\label{fig9}
\end{figure*}

\begin{figure}
\raggedleft
\includegraphics[width=1.0\linewidth]{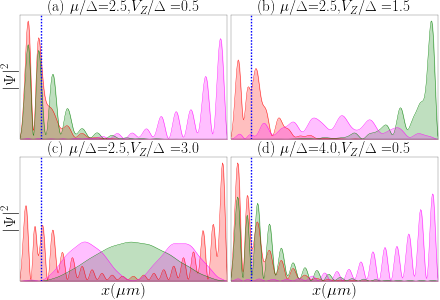}
\caption{(Color online) The spatial distribution of the low-energy state wave functions in Fig.~\ref{fig9} at various Zeeman fields for (a)--(c) $\mu=2.5\Delta$ and for (d) $\mu=4\Delta$. The first, second, and third lowest energy eigenstates are in red, green, and purple colors, respectively. The blue dotted lines mark the boundary between quantum dot and nanowire.}
\label{fig10}
\end{figure}

In all of the previous sections, we were focusing on pristine clean nanowires, i.e., nanowires with no quantum dot or chemical potential fluctuation or any kind of disorder. In such nanowires, an intrinsic ABS forms at each wire end at finite chemical potential in the topologically trivial regime. The corresponding tunneling conductance spectroscopy shows strong gap closure features below TQPT but no gap reopening features above TQPT. On the other hand, as has been already discussed thoroughly in the recent literature~\cite{Moore2016Majorana, Liu2017Andreev, Chiu2017Conductance,Prada2017Measuring,Clarke2017Experimentally,Liu2018Distinguishing}, if an external quantum dot or smooth confinement potential exists at one end of the nanowire, or there is chemical potential inhomogeneity inside the nanowire, a nontopological almost-zero-energy ABS can form, mimicking many features of MBS~\cite{Brouwer2011Topological, Chiu2017Conductance, Moore2016Majorana, Liu2017Andreev}. In this paper, we call such smooth potential-induced states extrinsic ABS, in contrast to intrinsic ABS for pristine clean nanowires being considered in the current work. The tunneling spectroscopy for extrinsic ABS also shows only gap closure features but no gap reopening features, although both features happen in the topologically trivial regime~\cite{Liu2017Andreev}. Essentially, all Andreev bound states produce strong gap closing type features in the conductance spectroscopy, independent of whether they are intrinsic or extrinsic.  The question we address in this section is what happens when both intrinsic and extrinsic bound states are present in the system. In this section, in order to study the intrinsic and extrinsic ABS together in a single hybrid structure, we consider nanowires of finite chemical potential in the presence of an external quantum dot. The Hamiltonian for the nanowire in the NS junction has exactly the same form as Eq.~\eqref{eq:H}, except that there is an additional non-SC quantum dot located between the lead and the clean nanowire. The schematic is shown in Fig.~\ref{figschQD}. The BdG Hamiltonian for the non-SC quantum dot is
\begin{equation}
H_\text{QD} = \left( -\frac{\hbar^2}{2m^*} \partial^2_x -i \alpha_R \partial_x \sigma_y +V(x)- \mu \right)\tau_z + V_Z\sigma_x.
\label{eq:Hqd}
\end{equation}
Here, the smooth confinement potential arising from the quantum dot is chosen to be $V(x)=V_D\cos{(\frac{3\pi x}{2L_{dot}})}$ with the potential amplitude being $V_D=0.8$~meV and the dot length being $L_{dot}=0.3~\mu$m.

The numerically calculated differential conductance (energy spectra) for the dot-nanowire hybrid structure is shown in the upper (lower) panel of Fig.~\ref{fig9}. Figure~\ref{fig9}(a) shows the conductance for a nanowire with zero chemical potential ($\mu=0$). When the chemical potential is zero, there is neither intrinsic nor extrinsic ABS in the nanowire, and thus the only conductance features inside the superconducting gap are the MBS-induced ZBCP above the TQPT($V_Z>\Delta$). Figures ~\ref{fig9}(b)--\ref{fig9}(d) show the conductance for dot-nanowire structures with finite chemical potential. As the Zeeman field increases from zero to the critical value ($V_{Zc} = \sqrt{\mu^2+\Delta^2}$), we can see a strong coherence peak marking the `gap closure' features but with no `gap reopening' features, which resembles the observations in pristine nanowires with finite chemical potential. 

We emphasize that such `gap closure' features in the dot-nanowire structures are due to extrinsic ABS which are mostly located inside the quantum dot. These extrinsic ABS in the dot-nanowire structures differ from intrinsic ABS in the pristine nanowire in three aspects. These three qualitative differences enable one to discern whether an actual experiment is observing the intrinsic ABS (with the ZBCP in the topological regime) or the extrinsic ABS (with the ZBCP in the trivial regime). First, extrinsic ABS-induced coherence peak occupies the whole trivial regime, as shown in Figs.~\ref{fig9}(b)--\ref{fig9}(d). While for pristine nanowire, intrinsic ABS-induced coherence peak will always merge into the particle-hole continuum at $V_{Zt} < V_{Zc}$ before TQPT. Second, extrinsic ABS-induced coherence peaks are at finite voltage below the SC gap at zero Zeeman field, while the peak in the pristine nanowire appears at the gap edge. This is because the corresponding wave function for the coherence peak is largely located inside the quantum dot and thus not fully proximitized by the parent SC. Third, the coherence peak at zero Zeeman field splits into two peaks when the Zeeman field is turned on in the extrinsic situation. One peak enters the bulk continuum and disappears at finite Zeeman field, while the other approaches the zero-bias voltage and transmutes into the strong ZBCP near TQPT. Such a peak splitting is associated with the Kramer's pair inside the external quantum dot. By contrast, in the pristine nanowire, the intrinsic ABS is always doubly degenerate even at finite Zeeman field in the long wire limit, with one being located at each wire end. Although we here use a specific form for the smooth potential $V(x)$, the numerical results obtained are generic, i.e., the qualitative features of the tunneling conductance still hold for other potential profiles as long as the potential is smooth. We put additional numerical results for quantum dots with other confinement potential profiles, such as exponential and Gaussian functions in Appendix B.

We show the corresponding energy spectra in the lower panels of Fig.~\ref{fig9}. Figure~\ref{fig9}(e) is the energy spectrum for dot-nanowire structure with zero chemical potential ($\mu=0$). In the topologically trivial regime, the two lowest eigenenergies (red and green lines) are inside the bulk continuum. Figures~\ref{fig9}(f)--\ref{fig9}(h) are the energy spectra for dot-nanowire structure with increasing finite chemical potential. Note that these two lowest eigenenergies are inside the SC gap and non-degenerate. These two eigenstates are localized dot-induced extrinsic ABS responsible for the strong coherence peaks observed in Figs.~\ref{fig9}(b)--\ref{fig9}(d). Such non-degeneracy among the lowest two eigenenergies is different from the pristine clean nanowire case. For the latter, the two intrinsic ABS are always nearly-degenerate for $0 < V_Z < V_{Zt}$. An interesting observation is that at small but finite Zeeman field ($V_Z < V_{Zc}$), the second and the third lowest eigenenergies anticross with each other. When the Zeeman field is larger than the anticrossing point, the corresponding tunneling conductance becomes weak [Figs.~\ref{fig9}(b)--\ref{fig9}(d)].

In order to clarify what is going on around the anticrossing point and the roles of intrinsic and extrinsic ABS, we further show the wave functions of the first three lowest eigenstates in Fig.~\ref{fig10}. Figures~\ref{fig10}(a)--\ref{fig10}(c) show the evolution of the three wave functions with increasing Zeeman field at fixed chemical potential $\mu=2.5\Delta$. In Fig.~\ref{fig10}(a), the Zeeman field is less than that for the anticrossing point. The first two eigenstates (red and green) are localized inside the external quantum dot on the left, while the third eigenstate (purple) is localized at the opposite end of the nanowire on the right. The first two eigenstates are extrinsic ABS responsible for the coherence peaks, because they are closer to the normal lead and thus more visible. The third eigenstate is intrinsic ABS which are invisible in the tunnel spectroscopy since they are much spatially further away from the lead. Figure~\ref{fig10}(b) shows the wave functions when the Zeeman field is larger than that of the crossing point but still smaller than the critical value ($V_{Zc}$). The lowest eigenstate is still the extrinsic ABS localized in the dot, while the second lowest eigenstate becomes intrinsic ABS localized at the far end of the nanowire, and the third lowest eigenstate becomes a bulk state. Therefore, in the tunnel conductance spectroscopy, we can only see a single coherence peak for the lowest eigenstate, as shown in Figs.~\ref{fig10}(b)--(d). When the Zeeman field is larger than the critical value ($V_Z>V_{Zc}$), the system enters the topological regime and a pair of localized MBS form at wire ends (red), the second and third eigenstates are both extended bulk states, as indicated by Fig.~\ref{fig10}(c). To show that the coexistence of extrinsic and intrinsic ABS is generic, we show in Fig.~\ref{fig10}(d) the wavefunctions at small Zeeman field for a different chemical potential $\mu=4\Delta$. Similar to Fig.~\ref{fig10}(a), the first two eigenstates are extrinsic ABS localized in the quantum dot, and the third state is intrinsic ABS localized at the far end of the nanowire.

\section{Conclusion}\label{sec:conclusion}
We report here the theoretical finding, supported by extensive numerical simulations, of a phenomenon in the extensively studied semiconductor-superconductor hybrid nanowire systems, which are of great current interest in the context of Majorana-based topological quantum computation.  We find that even in the ideal clean situation with no chemical potential fluctuations or quantum dot physics, the observed experimental SC gap closing feature with increasing magnetic field arises from the appearance of intrinsic ABS at the SC gap edges.  At zero chemical potential (the case extensively studied in the theoretical literature), these ABS do not exist, and the standard picture of the SC gap states being extended bulk band states apply, but in the generic situation of a finite chemical potential, a critical field $V_{Zt}$ emerges, and for $0<V_{Zt}<V_{Zc}$, where $V_{Zc}$ is the critical field for TQPT (where the trivial gap vanishes), the gap closing is associated with these localized ABS rather than extended bulk states.  We believe that the observed experimental gap closing in nanowires corresponds to the coming together of these ABS states, even if the system is ultraclean with no disorder.  Of course, it is already known that in disordered systems, chemical potential fluctuations lead to the existence of trivial ABS mimicking MBS~\cite{Brouwer2011Topological,Chiu2017Conductance,Liu2017Andreev,Moore2016Majorana}.  This work makes the situation even more interesting (and complicated) by showing that even in the pristine clean system, ABS exist almost all the way to the TQPT since for large generic $\mu$, $V_{Zt}$ approaches $V_{Zc}$ from below. The fact that the trivial gap below the TQPT in Majorana nanowires most likely arises from the existence of close-to-the-gap-edge generic  localized intrinsic ABS changes the understanding of the relevant physics for these systems since one can think of the TQPT as effectively transforming finite-energy in-gap ABS into zero-energy topological MBS with the conductance spectral weight carried mostly by the ABS (on the trivial side) and MBS (on the topological side). This ABS to MBS transmutation is understandable here since the ABS on the trivial side correspond to four degenerate MBSs, with two MBSs transmuting into zero-energy Majorana bound states localized at the ends of the wire and two becoming the gap-edge extended bulk states on the topological side. Note that the conductance spectral weight is concentrated entirely on the bound states on both sides and therefore, gap closing (opening) below (above) TQPT is visible (invisible). It should be emphasized that the actual TQPT always involves a true SC bulk gap closing since $V_{Zt}<V_{Zc}$, except that $V_{Zt}$ may be arbitrarily close to $V_{Zc}$ for large $\mu$ as can be seen in the phase diagram of Fig.~\ref{fig3}.

Our work, quite apart from being conceptually important, has immediate significance for nanowire MBS experiments.  In particular, large $\mu$ (which for all practical purposes boils down to $\mu>\Delta$, the induced proximity SC gap) should be generic in all experimental situations since $\Delta$ is typically small ($< 0.1$ meV).  Thus, it is likely that in all experimental systems $V_{Zt}\sim V_{Zc}$ within the experimental resolution.  It follows therefore that the magnetic field tuned gap closing feature observed experimentally in the best samples~\cite{Zhang2017Quantized, Deng2016Majorana,Nichele2017Scaling} \textit{always} manifests the closing of an ABS gap (i.e., coming together of localized bound states at the wire ends rather than extended bulk states), independent of whether the sample is clean (no chemical potential fluctuations or quantum dots) or not. Once the gap closes at $V_{Zc}$ ($\sim V_{Zt}$), however, the gap reopening states on the higher field side ($V_Z> V_{Zc}\sim V_{Zt}$) are extended bulk states since the ABS have metamorphosed into two zero-energy MBSs and one bulk state at $V_Z=V_{Zc}$ ($\sim V_{Zt}$). Thus, the physics of the semiconductor-superconductor hybrid nanowire at finite chemical potential is very different from the hypothetical (and physically irrelevant) zero chemical potential situation: For $0<V_Z<V_{Zt}\sim V_{Zc}$, the gap closing feature is the coming together of the ABS at $V_Z=V_{Zt}\sim V_{Zc}$ (in contrast to the $\mu=0$ case where the coming together of extended bulk states does not generate a clear gap closing feature in a tunneling conductance) whereas the gap reopening feature for $V_Z>V_{Zc}$ ($\sim V_{Zt}$) is the opening of bulk band states (as in the $\mu=0$ case). This finding explains a hitherto puzzling experimental feature observed universally.  Experimentally, the gap closing feature in the tunneling spectra is always prominent for $V_Z<V_{Zc}$, but the gap reopening feature is never prominent. This was challenging to understand until this work because the belief was that both gap closing ($V_Z<V_{Zc}$) and gap opening ($V_Z>V_{Zc}$) features arise from extended bulk states and, in fact, neither should be prominent in a tunneling experiment which only probes localized wave functions at the ends of the nanowire~\cite{Stanescu2012To}. Our current work, by contrast, establishes definitively that the low-field gap closing feature is generically a closing of an ABS gap, which should be prominent in a tunneling measurement which probes the local wave functions at the wire ends where these ABS are localized.  Thus, the mystery of the prominent gap closing feature in the tunneling measurements is solved by our work:  it arises from the generic existence of ABS for $V_Z<V_{Zc}\sim V_{Zt}$ at finite chemical potential. (We mention that it was pointed out in Ref.~\cite{Stanescu2012To}, based on numerical simulations, that the lack of a gap closing feature in the tunnel conductance may not be universal in short wires and in particular, for large $\mu$ it may be possible for gap closing to be visible, but the generic existence of ABS in the trivial phase, its transmutation to MBS, and its theoretical and experimental implications are findings of this work.) Equally importantly, our work also resolves the problem of experimental nonobservation of a gap reopening feature for $V_Z>V_{Zc}$ since the gap reopening states are extended bulk gap states for $V_Z>V_{Zc}$ -- the ABS for $V_Z<V_{Zt}\sim V_{Zc}$ have become the MBS for $V_Z>V_{Zc}\sim V_{Zt}$, with all above-gap states being extended bulk states which have little weight in tunneling measurements.

We mention that our current result on the existence of ABS in the SC gap of the nanowire below the TQPT has superficial formal similarity to the existence of excitons in semiconductor band gaps.  In both cases, the lowest energy eigenstates of the system (ABS in our case and excitons in semiconductors) are localized electron-hole bound states with the continuum bulk extended band states lying higher in energy above the gap.  In both cases, the spectral weight (or the local density of states) is carried dominantly by the bound state, leading to observable experimental consequences. Although very different in their physical origins, the analogy between our ABS and semiconductor excitons provides a physical intuition for the ABS dominating the trivial phase conductance in the nanowire tunneling measurements.  Of course, the physics of TQPT and the transmutation of the ABS into MBS above the TQPT are qualitative physics present only in the semiconductor-superconductor hybrid structures. Also, the fact that one needs spin-orbit coupling, Zeeman splitting, and superconductivity together to produce the Andreev bound states in the superconducting gap of the nanowire is specific to the current hybrid system -- by contrast, excitons in the semiconducting energy gap form because of Coulomb interaction between electrons and holes whereas Coulomb coupling does not play a role in the formation of ABS in the current system.

Our work does not, however, resolve the serious issue of whether the experimentally observed ZBCP arises from topological MBS or trivial ABS produced by chemical potential fluctuations, since our work does not affect at all the $V_Z>V_{Zc}$ topological side of Majorana physics. All we can say is that the low-field experimental gap closing features arise invariably from ABS before the putative TQPT, either from chemical potential fluctuations as in the existing work~\cite{Liu2017Andreev} or from intrinsic finite chemical potential effects as in this work.  Our work obviously does not rule out the possibility of the experimentally observed ZBCP arising from almost-zero accidental extrinsic ABS, but establishes that it is possible for low-field intrinsic ABS to transform into high-field MBS in clean samples with no chemical potential fluctuations. Our work shows that the low-field ($V_Z<V_{Zc}$) side of the SC spectrum is dominated by ABS generically, independent of chemical potential fluctuations or quantum dots being present or not.  Thus, the fact that there is a strong tunneling conductance feature associated with the gap closing for $V_Z<V_{Zc}$ has no significance with respect to the existence or not of MBS on the high-field side ($V_Z>V_{Zc}$). The existence of ABS on the low-field side neither rules out nor reinforces the possible existence of MBS on the high field side. The fact that experiments in the cleanest samples observe strong gap closing features and no gap opening features in the tunnel conductance cannot by itself be construed as indicating the existence of trivial accidental extrinsic ABS in the system.  Our work establishes this to be the generic behavior on the trivial side at finite chemical potential.  The key problem here is of course that theoretically we know the precise location of the TQPT whereas experimentally $V_{Zc}$ is not known, and hence it is difficult to know whether a ZBCP arises on the topological side ($V_Z>V_{Zc}$) from MBS or is simply an accidental feature of trivial ABS sticking to almost zero energy.

In this context, we have also discussed, providing extensive results, the interplay between the intrinsic ABS and the extrinsic ABS using a dot-nanowire hybrid structure where the quantum dot induces extrinsic ABS.  We find that the most significant feature distinguishing the two is whether the starting ABS energies at zero Zeeman field are at the gap edge (``intrinsic") or well inside the gap (``extrinsic").  Thus, experimentally one should be able to tell whether the gap closing feature is associated with intrinsic or extrinsic ABS simply by reading off the associated effective energy gap at zero field.  Also, the intrinsic ABS disappear at $V_{Zt} < V_{Zc}$, i.e., before the ZBCP forms whereas the extrinsic ABS themselves stick to zero energy, enabling another distinction between them.  We provide a detailed discussion on the differences between intrinsic and extrinsic ABS in Sec.~\ref{sec:QD} and Appendix B.

Finally, we mention that although our work is specifically for the semiconductor-superconductor nanowire systems, the physics we describe is generic and should apply to all proximity-induced SC systems where spin-orbit coupling and Zeeman splitting are operational, and we believe that the trivial side of TQPT in such systems is always dominated by ABS gap closing rather than a bulk gap closing in the generic finite chemical potential situation.

\section*{acknowledge}
This work is supported by Laboratory for Physical Sciences and Microsoft. Y.H. acknowledges the funding from China Scholarship Council. We also acknowledge the support of the University of Maryland supercomputing resources (http://hpcc.umd.edu/).

\bibliography{BibMajorana}
\onecolumngrid
\vspace{1cm}
\begin{center}
{\bf\large Appendix A}
\end{center}
\vspace{0.5cm}

\setcounter{secnumdepth}{3}
\setcounter{equation}{0}
\setcounter{figure}{0}
\renewcommand{\theequation}{S-\arabic{equation}}
\renewcommand{\thefigure}{A\arabic{figure}}
\renewcommand\figurename{Figure}
\renewcommand\tablename{Supplementary Table}
\newcommand\Scite[1]{[S\citealp{#1}]}
\makeatletter \renewcommand\@biblabel[1]{[S#1]} \makeatother

\begin{figure*}
\raggedleft
\includegraphics[width=1.0\textwidth]{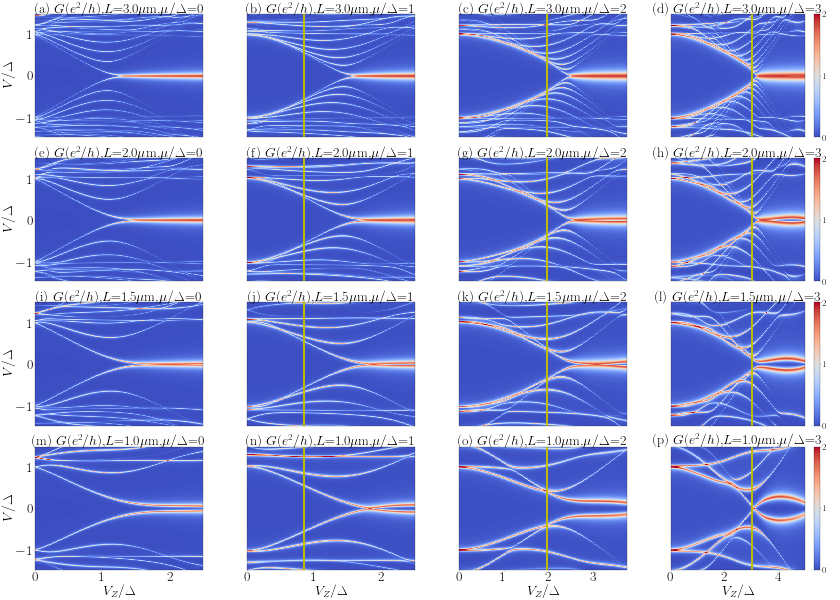}
\caption{(Color online) Numerical calculated differential tunneling conductance with $\alpha_R=0.5~$eV$\angstrom$ in short pristine Majorana nanowires as a function of Zeeman field for various chemical potentials and wire lengths.
The panels in the four columns show conductance with increasing chemical potentials (from left to right): $\mu/\Delta=0, 1, 2, 3$.
The panels in the four rows show conductance with different wire lengths (from top to bottom): $L\cong3, 2, 1.5, 1\mu$m.
The yellow vertical line marks the $V_{Zt}$ calculated by Eq.\eqref{Eq:Vzt} in infinite-long-wire limit.
Similar to the conductance for wire length $L\cong20\mu$m in Fig.~\ref{fig1}, there is a strong `gap closure' feature below $V_{Zt}$ in the short nanowires. On the other hand, the splitting of the conductance peak in trivial region is obvious [e.g., (h) and (l)] since the ABS in the two wire ends overlap with each other in the middle region in a the short wire.
}
\label{Afig1}
\end{figure*}

\begin{figure*}
\raggedleft
\includegraphics[width=1.0\textwidth]{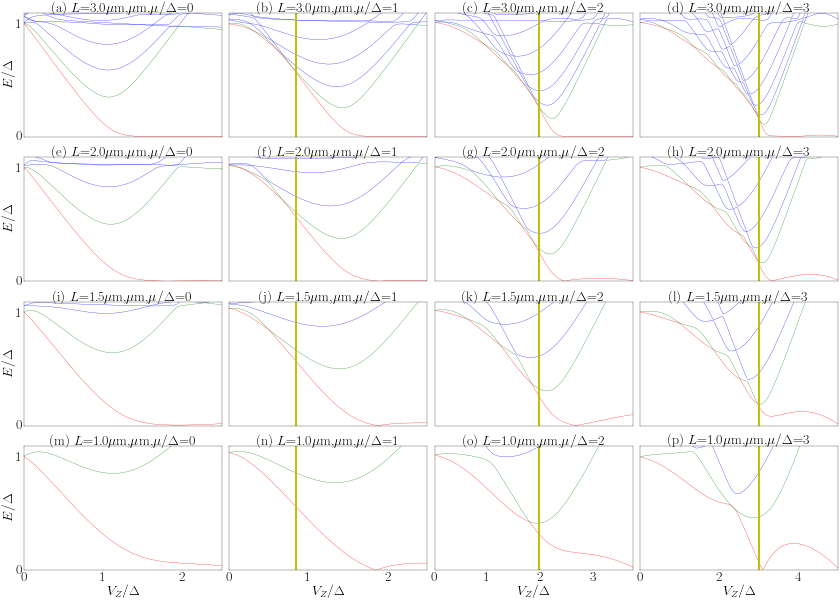}
\caption{(Color online)  The low-energy spectra corresponding to the conductance in Fig.~\ref{Afig1} for $\alpha_R=0.5~$eV$\angstrom$. While the blue lines denote the bulk states, the red and green lines are associated with the first and second lowest-energy eigenstates, respectively. The yellow vertical line marks the $V_{Zt}$ calculated by Eq.~\eqref{Eq:Vzt} in infinite-long-wire limit. The energies of the two lowest eigenstates below $V_{Zt}$ split and oscillate. This is the result of the overlapping of ABS in a finite-length wire. }
\label{Afig2}
\end{figure*}

\begin{figure*}
\raggedleft
\includegraphics[width=1.0\textwidth]{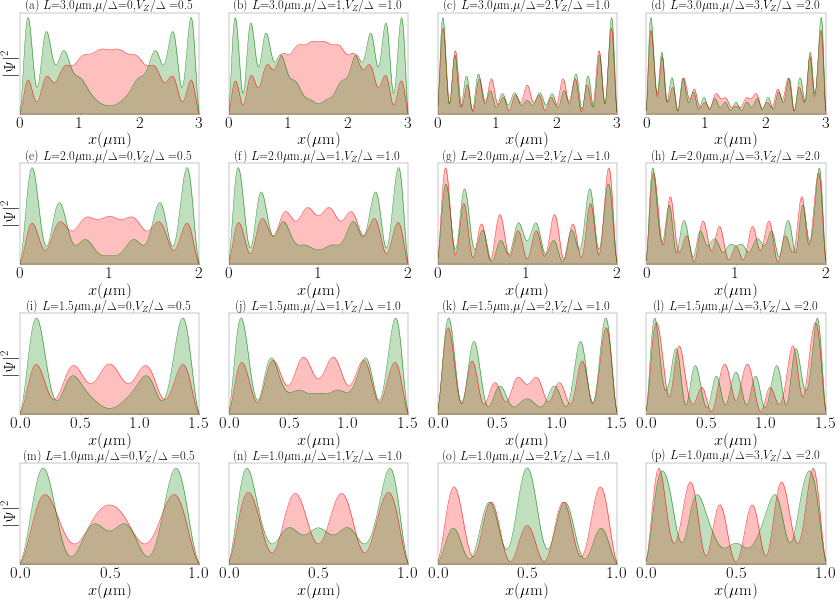}
\caption{(Color online) The wave functions of the lowest two eigenstates below TQPT corresponding to Figs.~\ref{Afig1} and \ref{Afig2} for spin-orbit coupling $\alpha_R=0.5~$eV$\angstrom$. The first (second) lowest energy eigenstates are in red (green) color. The four rows correspond to different lengths of the nanowire, i.e. $L = 3, 2, 1.5, 1\mu$m. In the first row, when chemical potential is small, (a) and (b) show extended bulk states inside the nanowire. As the chemical potential increases, (c)(d) show two ABS localized at two wire ends. For each column, as the wire length decreases, the bulk states remain extended [(e), (f), (i), (j), (m), (n)], while the ABS cross over from localized states in the long-wire limit to extended states in the short-wire limit (k,l,o,p).
}
\label{Afig3}
\end{figure*}

\begin{figure*}
\raggedleft
\includegraphics[width=1.0\textwidth]{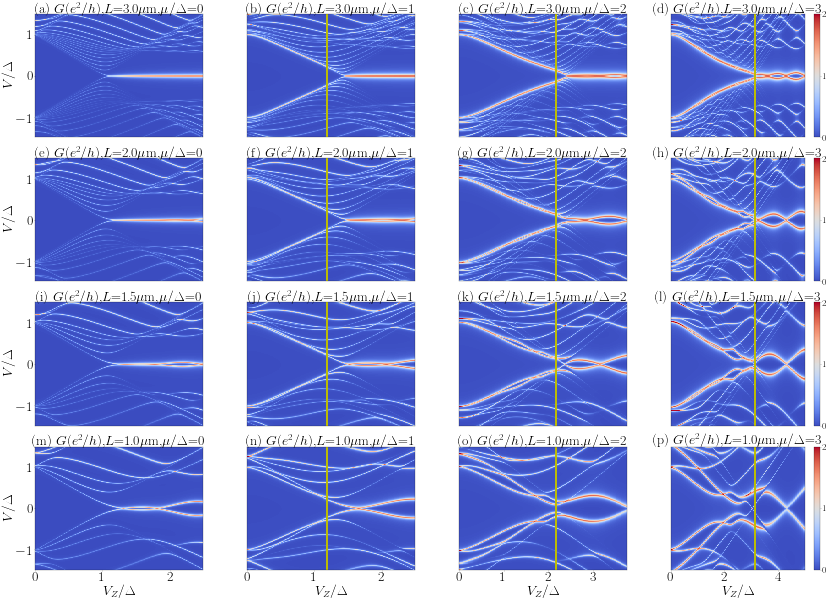}
\caption{(Color online) Numerical calculated differential tunneling conductance with $\alpha_R=0.2~$eV$\angstrom$ in short pristine Majorana nanowires as a function of Zeeman field for various chemical potentials and wire lengths.
The panels in the four columns show conductance with increasing chemical potentials (from left to right): $\mu/\Delta=0, 1, 2, 3$.
The panels in the four rows show conductance with different wire lengths (from top to bottom): $L\cong3, 2, 1.5, 1\mu$m.
The yellow vertical line marks the $V_{Zt}$ calculated by Eq.\eqref{Eq:Vzt} in infinite-long-wire limit.
Compared with the conductance for spin-orbit coupling $\alpha_R=0.5~$eV$\angstrom$ in Fig.~\ref{Afig1}, the `gap closure' feature below $V_{Zt}$ is stronger and its oscillation is weaker at $\alpha_R=0.2~$eV$\AA$. This is due to the short ABS localization length at $\alpha_R=0.2~$eV$\AA$ as indicated in Fig.~\ref{fig6}.
}
\label{Afig4}
\end{figure*}

\begin{figure*}
\raggedleft
\includegraphics[width=1.0\textwidth]{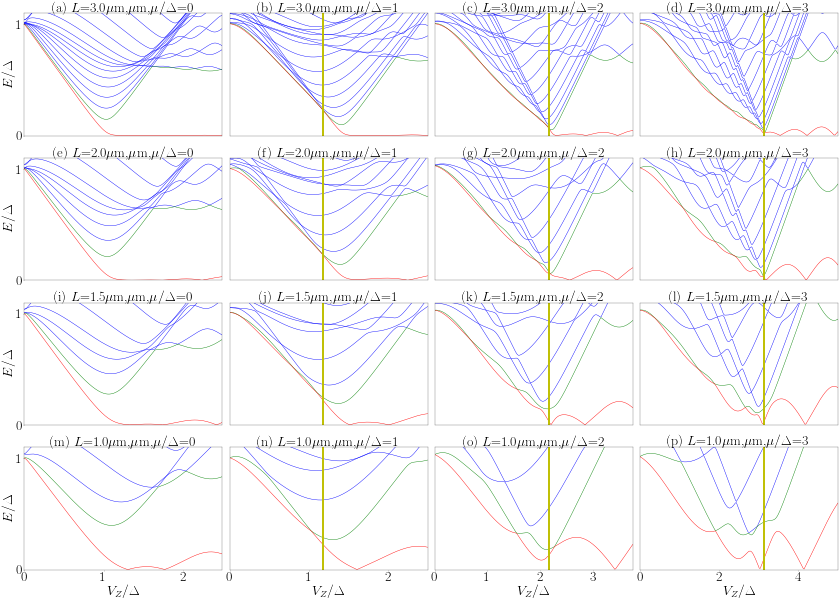}
\caption{
(Color online) The low-energy spectra corresponding to the conductance in Fig.~\ref{Afig4} for $\alpha_R=0.2~$eV$\angstrom$. While the blue lines denote the bulk states, the red and green lines are associated with the first and second lowest energy eigenstates. The yellow vertical line marks the $V_{Zt}$ calculated by Eq.\eqref{Eq:Vzt} in infinite-long-wire limit. In comparison with Fig.~\ref{Afig2} for $\alpha_R=0.5~$eV$\angstrom$, the energy splitting induced by shorter ABS localization length is weaker at $\alpha_R=0.2~$eV$\angstrom$. }
\label{Afig5}
\end{figure*}

\begin{figure*}
\raggedleft
\includegraphics[width=1.0\textwidth]{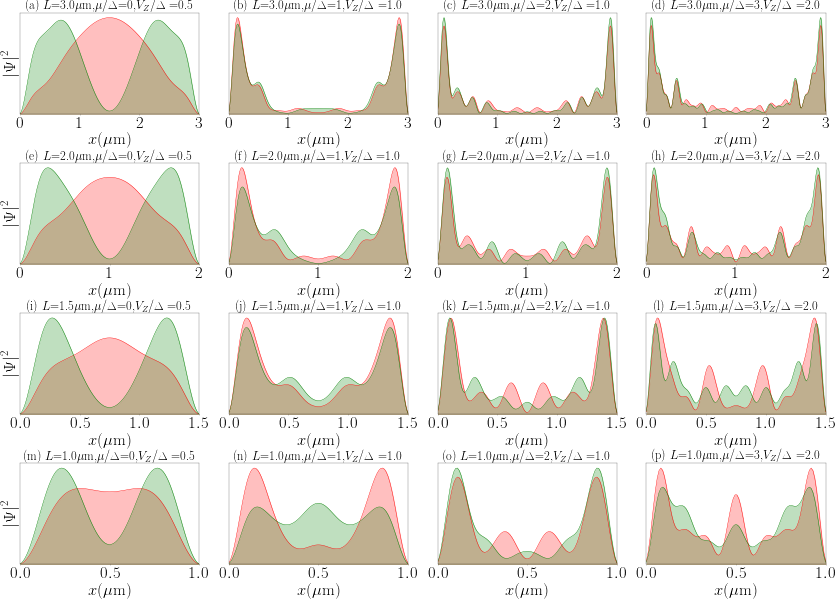}
\caption{(Color online) The wave functions of the the lowest two eigenstates below TQPT in Figs.~\ref{Afig4} and \ref{Afig5} for $\alpha_R=0.2~$eV$\angstrom$.  The first (second) lowest-energy eigenstates are in red (green) color. The four rows correspond to different lengths of the nanowire, i.e. $L = 3,2,1.5,1\mu$m.
(a) Show extended bulk states, while (b), (c), (d) show two degenerate ABSs localized at two wire ends.
In comparison with Fig.~\ref{Afig3} for $\alpha_R=0.5~$eV$\angstrom$, the overlapping of ABS for $\alpha_R=0.2~$eV$\angstrom$ is smaller [(c), (d), (g), (h)].}
\label{Afig6}
\end{figure*}

\begin{figure*}
\raggedleft
\includegraphics[width=1.0\textwidth]{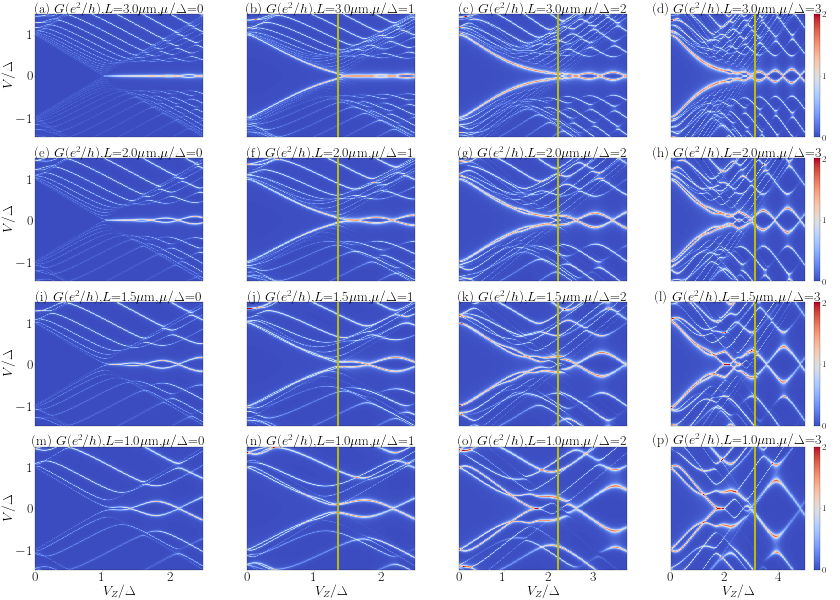}
\caption{(Color online) Numerical calculated differential tunneling conductance with spin-orbit coupling $\alpha_R=0.1~$eV$\angstrom$ in short pristine Majorana nanowires as a function of Zeeman field for various chemical potentials and wire lengths.
The panels in the four columns show conductance with increasing chemical potentials (from left to right): $\mu/\Delta=0,1,2,3$.
The panels in the four rows show conductance with different wire lengths (from top to bottom): $L\cong3,2,1.5,1\mu$m.
The yellow vertical line marks the $V_{Zt}$ calculated by Eq.\eqref{Eq:Vzt} in infinite-long wire limit.
The `gap closure' feature below $V_{Zt}$ is still strong for $\alpha_R=0.1~$eV$\AA$ in comparison with the conductance for $\alpha_R=0.5~$eV$\angstrom$ in Fig.~\ref{Afig1} and $\alpha_R=0.2~$eV$\angstrom$ in Fig.~\ref{Afig4}.  }
\label{Afig7}
\end{figure*}

\begin{figure*}
\raggedleft
\includegraphics[width=1.0\textwidth]{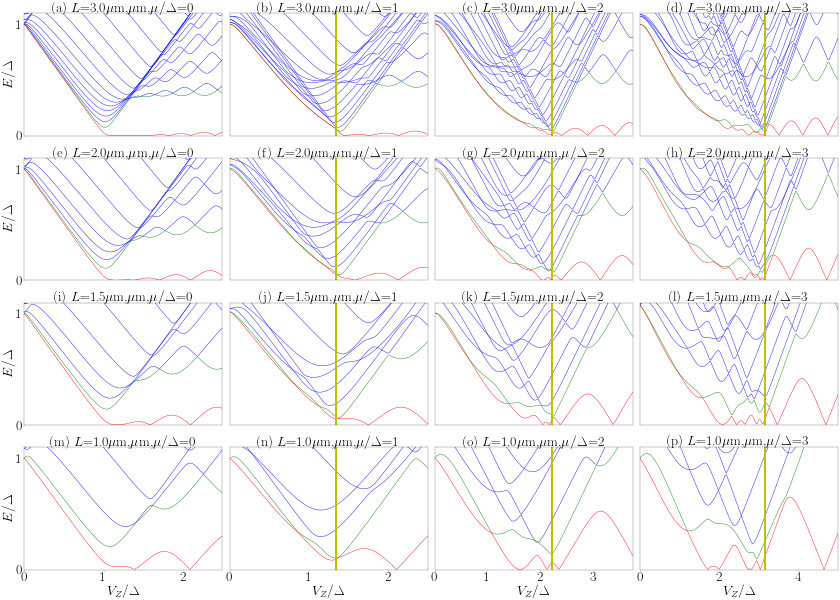}
\caption{
(Color online) The low-energy spectra corresponding to the conductance in Fig.~\ref{Afig7} for spin-orbit coupling $\alpha_R=0.1~$eV$\angstrom$. While the blue lines denote the bulk states, the red and green lines are associated with the first and second lowest-energy eigenstates. The yellow vertical line marks the $V_{Zt}$ calculated by Eq.\eqref{Eq:Vzt} in infinite-long-wire limit. }
\label{Afig8}
\end{figure*}

\begin{figure*}
\raggedleft
\includegraphics[width=1.0\textwidth]{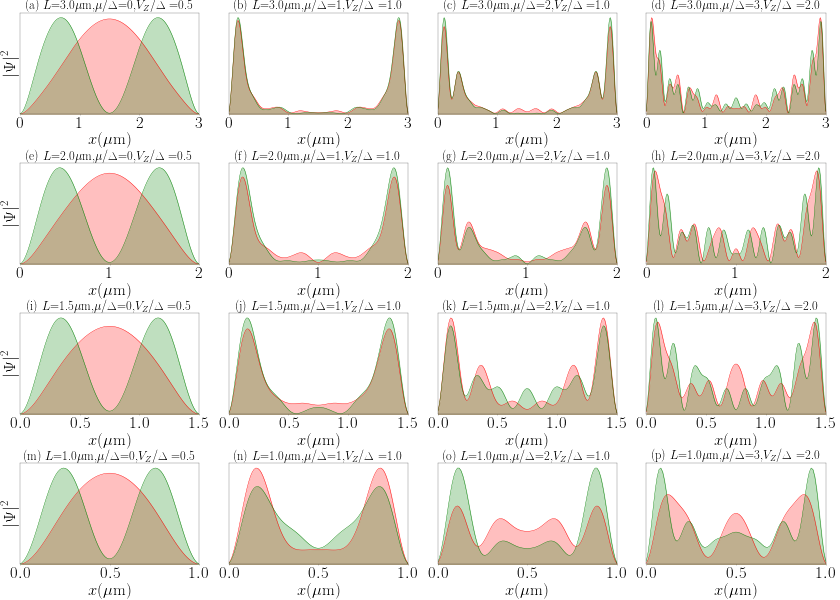}
\caption{(Color online) The wave functions of the lowest two eigenstates below TQPT in Figs.~\ref{Afig7} and \ref{Afig8} for $\alpha_R=0.1~$eV$\angstrom$.  The first (second) lowest energy eigenstates are in red (green) color.
The four rows correspond to different lengths of the nanowire, i.e. $L = 3, 2, 1.5, 1 \mu$m. (a) shows extended bulk states, while (b), (c), (d) show two degenerate ABSs localized at two wire ends.
}
\label{Afig9}
\end{figure*}

\begin{figure*}
\centering
\raggedleft
\includegraphics[width=1.0\textwidth]{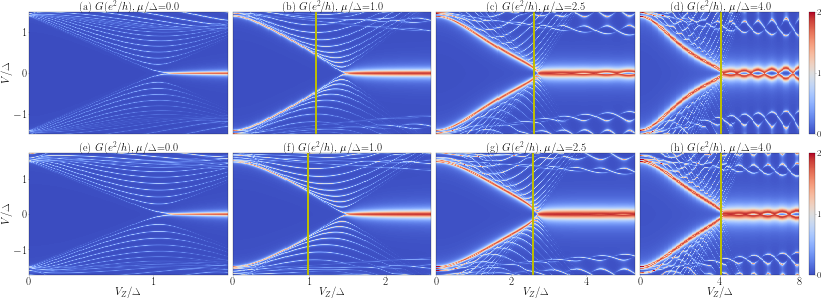}
\caption{(Color online) Numerical calculated differential tunneling conductance in pristine Majorana nanowires with length $L\cong3\mu$m as a function of Zeeman field for various chemical potentials. The yellow vertical line marks the $V_{Zt}$ calculated by Eq.~\eqref{Eq:Vzt} in infinite-long wire limit.
In the upper panels, (a)--(d), the spin-orbit coupling strength and SC gap are chosen to be $\alpha_R=0.3~$eV$\angstrom$ and $\Delta=0.28$~meV for a small ABS localization length.
In the lower panels, (e)--(h), the spin-orbit-coupling strength is $\alpha_R=0.4~$eV$\angstrom$ and SC gap is $\Delta=0.3$~meV for another small ABS localization length. Both the ABS gap closure features are strong for the two small ABS localization lengths.}
\label{Afig10}
\end{figure*}

\begin{figure*}
\raggedleft
\includegraphics[width=1.0\textwidth]{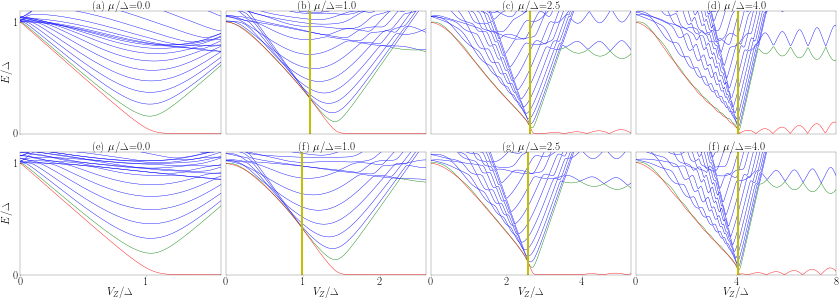}
\caption{(Color online) The low-energy spectra corresponding to the conductance in Fig.~\ref{Afig10}. While the blue lines denote the bulk states, the red and green lines are associated with the first and second lowest energy eigenstates. The energy splittings of the ABS are very weak due to the small ABS localization lengths. }
\label{Afig11}
\end{figure*}

\begin{figure*}
\raggedleft
\includegraphics[width=1.0\textwidth]{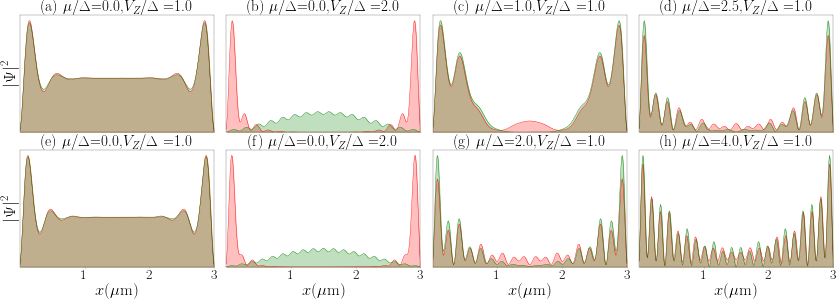}
\caption{(Color online) The wavefunctions of the lowest two eigenstates in Figs.~\ref{Afig10} and \ref{Afig11}. The first (second) lowest-energy eigenstates are in red (green) colors. (a), (e) show extended bulk states, and (b), (f) show MBS, while (c), (d) and (g), (h) show ABS localized at two wire ends. }
\label{Afig12}
\end{figure*}

In this Appendix we provide several sets of calculated conductance, energy spectra, and wave functions for the Majorana nanowire using different values of system parameters in order to demonstrate the generic existence of ABS in the nontopological phase in pristine nanowires for finite chemical potential and Zeeman splitting (see Figs.~\ref{Afig1}--\ref{Afig12}).  These calculations follow the same techniques as described in the main text of the paper.




\clearpage

\vspace{1cm}
\begin{center}
{\bf\large Appendix B}
\end{center}
\vspace{0.5cm}

\setcounter{secnumdepth}{3}
\setcounter{equation}{0}
\setcounter{figure}{0}
\renewcommand{\theequation}{B-\arabic{equation}}
\renewcommand{\thefigure}{B\arabic{figure}}
\renewcommand\figurename{Figure}
\renewcommand\tablename{Supplementary Table}
\newcommand\Bcite[1]{[B\citealp{#1}]}
\makeatletter \renewcommand\@biblabel[1]{[B#1]} \makeatother
In this Appendix we provide several sets of calculated conductance, energy spectra, and wave functions for the hybrid quantum dot--nanowire--superconductor using different values of system parameters (see Figs.~\ref{Bfig1}--\ref{Bfig6}).  These calculations follow the same techniques as described in the Sec.~\ref{sec:QD} of the main text .

\begin{figure*}[htbp]
\raggedleft
\includegraphics[width=1.0\textwidth]{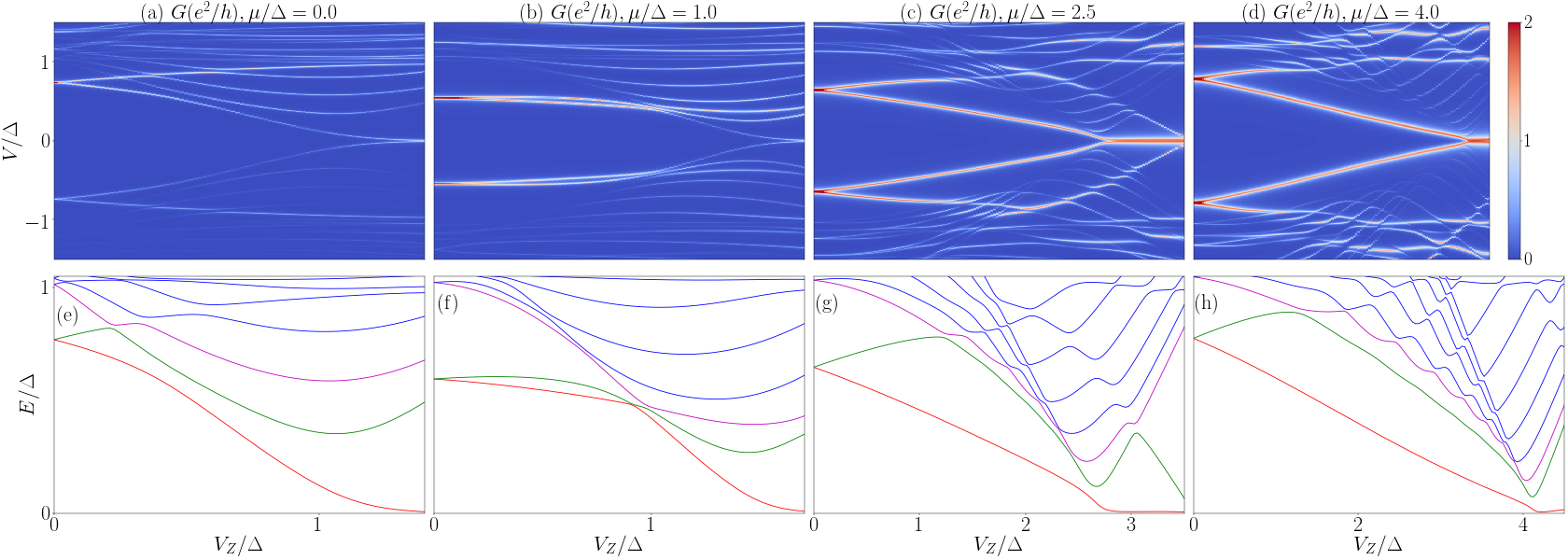}
\caption{(Color online) Upper panels: numerical calculated differential tunneling conductance with $\alpha_R=0.5~$eV$\angstrom$ through the quantum dot-nanowire-superconductor hybrid system as a function of Zeeman field for various chemical potentials. The wire length is $3\mu$m and the quantum dot is in a cosine function as $V(x)=V_D\cos({\frac{3\pi x}{2L_{dot}}})$
 with $V_D=0.8$~meV and $L_{dot}=0.3\mu$m.
Lower panels: the corresponding low-energy spectra as a function of Zeeman field. The red, green, and purple lines are associated with the first, second and third lowest energy eigenstates, while the blue lines denote the bulk states.
}
\label{Bfig1}
\end{figure*}

\begin{figure*}[htbp]
\raggedleft
\includegraphics[width=1.0\textwidth]{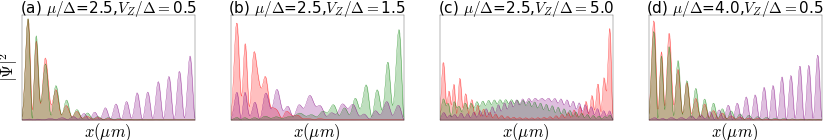}
\caption{(Color online)  The wave-functions corresponding to the conductance and energy spectra in Fig.~\ref{Bfig1} for $\alpha_R=0.5~$eV$\angstrom$.  The first, second, and third lowest-energy eigenstates are in red, green, and purple colors, respectively.}
\label{Bfig2}
\end{figure*}
\begin{figure*}[htbp]
\raggedleft
\includegraphics[width=1.0\textwidth]{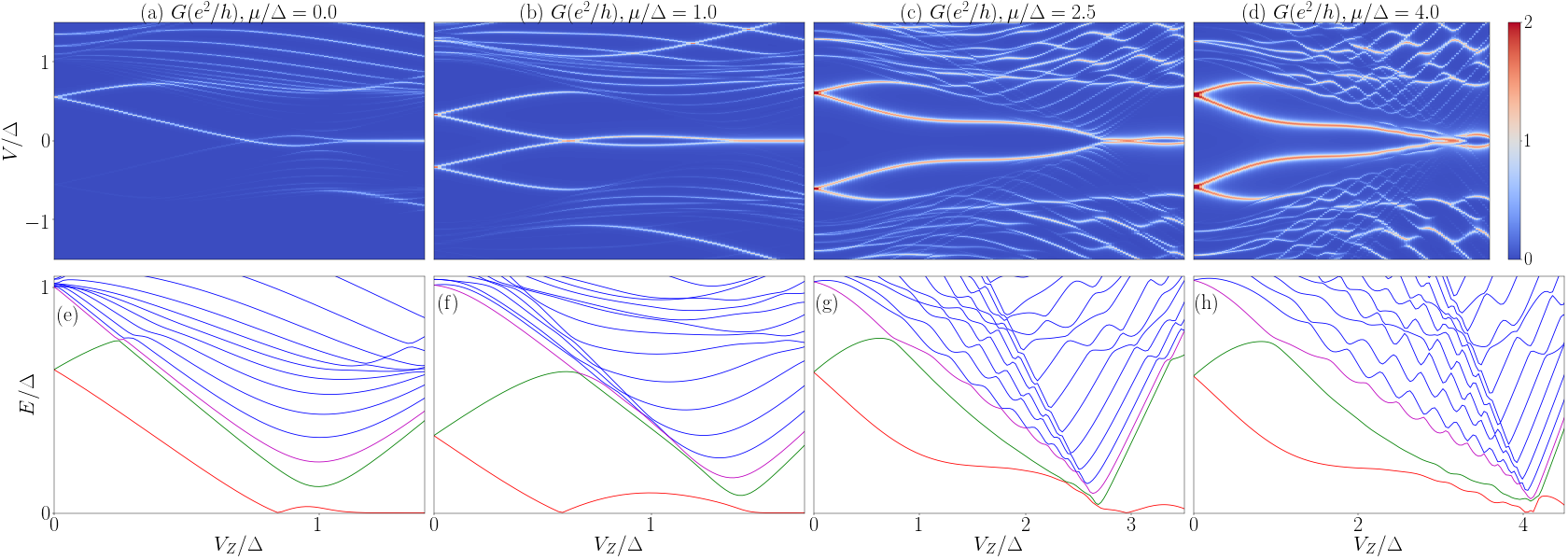}
\caption{(Color online) Upper panels: numerical calculated differential tunneling conductance through the quantum dot--nanowire--superconductor hybrid system as a function of Zeeman field for various chemical potentials with the quantum dot in an exponential function as $V(x)=V_D\exp(-5x/L_{dot})$.
Lower panels: the corresponding low-energy spectra as a function of Zeeman field. The red, green and purple lines are associated with the first, second and third lowest energy eigenstates, while the blue lines denote the bulk states.  The quantum dot strength is $V_D=0.8$~meV and the length is $L_{dot}=0.3\mu$m. The wire length is $L=3\mu$m and spin-orbit coupling $\alpha_R=0.2~$eV$\angstrom$.
}
\label{Bfig3}
\end{figure*}
\begin{figure*}[htbp]
\raggedleft
\includegraphics[width=1.0\textwidth]{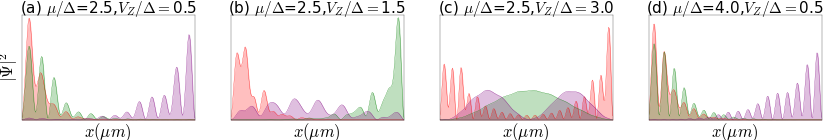}
\caption{(Color online)  The wave functions corresponding to the conductance and energy spectra in Fig.~\ref{Bfig3} for the quantum dot in an exponential function as $V(x)=V_D\exp(-5x/L_{dot})$.  The first, second, and third lowest energy eigenstates are in red, green, and purple colors, respectively.
}
\label{Bfig4}
\end{figure*}

\begin{figure*}[htbp]
\raggedleft
\includegraphics[width=1.0\textwidth]{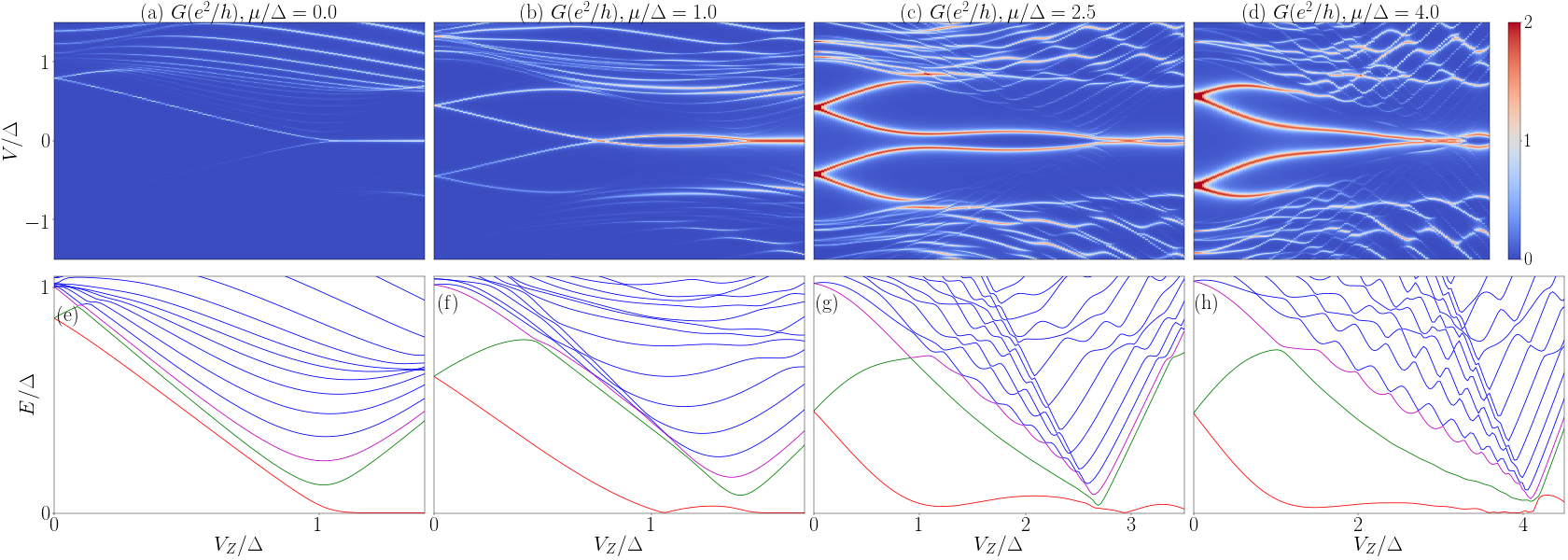}
\caption{(Color online) Upper panels: numerical calculated differential tunneling conductance through the quantum dot-nanowire-superconductor hybrid system as a function of Zeeman field for various chemical potentials with the quantum dot in a Gaussian function as $V(x)=V_D\exp[-\frac{(x-L_{dot}/2)^2}{2\sigma^2}]$.
Lower panels: the corresponding low-energy spectra as a function of Zeeman field. The red, green and purple lines are associated with the first, second and third lowest energy eigenstates, while the blue lines denote the bulk states.  The quantum dot strength is $V_D=0.8$~meV, the length is $L_{dot}=0.3\mu$m and the smoothness is $\sigma=0.03\mu$m. The wire length is $L=3\mu$m and spin-orbit coupling $\alpha_R=0.2~$eV$\angstrom$.
}
\label{Bfig5}
\end{figure*}
\begin{figure*}[htbp]
\raggedleft
\includegraphics[width=1.0\textwidth]{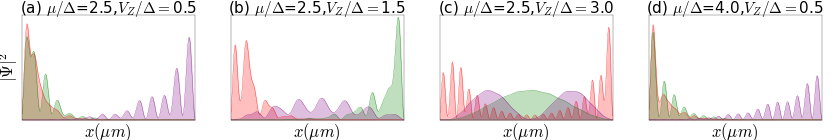}
\caption{(Color online)  The wave functions corresponding to the conductance and energy spectra in Fig.~\ref{Bfig5} for the quantum dot in a Gaussian function as $V(x)=V_D\exp[-\frac{(x-L_{dot}/2)^2}{2\sigma^2}]$.  The first, second, and third lowest energy eigenstates are in red, green, and purple colors, respectively.
}
\label{Bfig6}
\end{figure*}

\end{document}